\documentclass[a4paper,fleqn,usenatbib,useAMS]{mnras}
\usepackage[english]{babel}
\usepackage[T1]{fontenc}
\usepackage{amsmath}
\usepackage{amssymb}
\usepackage{booktabs}
\usepackage{graphicx,subfig}
\usepackage{color}
\usepackage{epstopdf}
\usepackage{comment}
\usepackage{caption}
\usepackage{natbib}
\usepackage{multirow}
\usepackage{subfig}
\usepackage[flushleft]{threeparttable}

\title[Prospects for detection of detached double white dwarf binaries  with Gaia, LSST and LISA]{Prospects for detection of detached double white dwarf binaries  with Gaia, LSST and LISA}
\author[V. Korol et al.]{Valeriya Korol$^{1}$\thanks{E-mail:korol@strw.leidenuniv.nl}, 
Elena M. Rossi$^1$, Paul J. Groot$^2$, Gijs Nelemans$^{2,3}$, Silvia Toonen$^{1,4}$, 
\newauthor Anthony G.A. Brown$^1$ \\
\\$^1$Leiden Observatory, Leiden University, PO Box 9513, 2300 RA, Leiden, the Netherlands
\\$^2$Department of Astrophysics, IMAPP, Radboud University, PO Box 9010, 6500 GL, Nijmegen, the Netherlands
\\$^3$Institute for Astronomy, KU Leuven, Celestijnenlaan 200D, 3001 Leuven, Belgium
\\$^4$Astronomical Institute Anton Pannekoek, University of Amsterdam, PO Box 94249, 1090 GE, Amsterdam, the Netherlands}
\date{March 1, 2017 - submitted to MNRAS}

\pagerange{\pageref{firstpage}--\pageref{lastpage}} \pubyear{2016}

\begin{document}
\maketitle
\label{firstpage}

\begin{abstract}

Double white dwarf (DWD) binaries are expected to be very common in the Milky Way, but their intrinsic faintness challenges the detection of these systems. 
Currently, only a few tens of {\it detached} DWDs are know. 
Such systems offer the best chance of extracting the physical properties that would allow us to address a wealth of outstanding questions ranging from the nature of white dwarfs, over stellar and binary evolution to mapping the Galaxy.
In this paper we explore the prospects for detections of ultra-compact (with binary separations of a few solar radii or less) detached DWDs in: 1) optical radiation with Gaia and the LSST and 2) gravitational wave radiation with LISA.
We show that Gaia, LSST and LISA have the potential to detect respectively around a few hundreds, a thousand, and 25 thousand DWD systems.
Moreover, Gaia and LSST data will extend by respectively a factor of two and seven the guaranteed sample of binaries detected in electromagnetic and gravitational wave radiation, opening  the era of multi-messenger astronomy for these sources.

\end{abstract}


%%%%%%%%%%%%%%%%%%%%%%%%%%%%%%%%%%%%%%%%%%%%%%%%%%%%%%%%%%%%%%
%%%%%%%%%%%%%%%%%%%%%%%%%%%%%%%%%%%%%%%%%%%%%%%%%%%%%%%%%%%%%%

\section{Introduction} 

On the basis of our theoretical understanding of stellar and binary evolution, systems of two white dwarfs in a close binary were predicted since 1980s (thereafter double white dwarf (DWD) binaries) \citep{Tutukov1981, Iben1984b, Webbink1984, Tutukov1988, Iben1997, Han1998,Nelemans2000,Nelemans2001a,Toonen2012}. 
However, due to their intrinsic faintness the first detection came only a decade later in 1988 \citep{Saffer1988}. 
The current census counts a few tens of DWDs discovered by spectroscopic and variability surveys such as the SPY (ESO SN Ia Progenitor) survey \citep[e.g.][]{SPY}, the ELM (Extremely Low Mass WDs) survey \citep[e.g.][]{ELM}, 
and studies by \cite{Marsh1995,MDD1995,MaxtedandMarsh1999} and \cite{Badenes2009}.
Still, these represents only a tiny fraction of DWD binaries predicted in numerical simulations (Toonen et al. 2017, submitted).

Substantial progress in the detection of these sources is expected with optical wide surveys such as Gaia \citep{Gaia} and the Large Synoptic Survey Telescope (LSST) \citep{LSST}, and in gravitational waves with the Laser Interferometer Space Antenna (LISA) mission \citep[e.g.][]{Amaro-Seoane2012}.
All three instruments will be sensitive to short period ($P$ < a few days) binaries \citep[e.g.,][]{Prsa2011, Eyer2012, Nelemans2013, Carrasco2014} and will provide a large sample of new ultra-compact DWDs that are interesting for several reasons.
First, compact DWDs are systems that experienced at least two phases of mass transfer, and thus provide a good test for binary evolution models, 
and, in particular, for our understanding of mass transfer and the common envelope (CE) phase.
Second, DWDs are the plausible progenitors to a wide range of interesting systems:
type Ia supernovae \citep{Iben1984a,Webbink1984}, that are used as cosmological distance indicators \citep[e.g.][]{Riess1998, Perlmutter1999}, AM CVn systems \citep[e.g.][]{Nelemans2001a,Marsh2004, Solheim2010} and type .Ia supernovae \citep{Bildsten2007}.
In addition, it is believed that the merger of two WDs can produce rare stars such as massive WDs (or even an isolated neutron star),  subdwarf-O and R Corona Borealis stars \citep{Webbink1984}.
Third, DWDs represent guaranteed sources for the LISA mission, and will dominate the low frequency gravitational wave band from mHz to a few Hz \citep[e.g.][]{Evans1987,Lipunov1987,Hils1990,Nelemans2001GW,Marsh2011}.
Finally, detached DWD binaries with orbital periods in the range from one hour to a few minutes are particularly suitable for studying the physics of tides, a phenomenon directly related to the WD internal properties.
The study of the reaction of the stellar internal structure to tidal forces may give us important information, for example, on WD viscosity and its origin, that will complete our knowledge on WD interior matter \citep{Piro2011,Fuller2012,DOR2014, McKernan2016}.

In this paper we compute the size of a sample of Galactic ultra-compact detached DWD binaries that could be obtained with future facilities in the next two decades.
In particular, we predict the size \citep[likewise][]{Cooray2004,Littenberg2013,Shah2013} and properties of the sample that will be observed in both electromagnetic (EM) and gravitational wave (GW) radiation by Gaia, LSST and LISA:
despite the widespread expectation that those instruments will represent major step forwards, quantitative predictions have never been published.
We characterize the physical properties of these samples and compare them to current data.

The paper is organized as follows.
In Section 2 we will describe the method we use to simulate the Galactic population of DWDs.
In Section 3 we will estimate how many binaries can be detected with Gaia and LSST as eclipsing sources.
In Section 4 we will focus on the GW emission from these sources and we assess the prospects for detections by the upcoming LISA mission. 
In Section 5 we will present and characterize the sample of DWDs detectable trough EM and GW radiation. 
Finally, we will discuss our results and possible synergies between GW and EM data.
 
%%%%%%%%%%%%%%%%%%%%%%%%%%%%%%%%%%%%%%%%%%%%%%%%%%%%%%%%%%%%%%
%%%%%%%%%%%%%%%%%%%%%%%%%%%%%%%%%%%%%%%%%%%%%%%%%%%%%%%%%%%%%%

\section{Simulated DWD population}

To obtain a model sample of the Galactic DWD population we use the binary population synthesis code SeBa, developed by \citet[][for updateds see \citealt{Nelemans2001a}, \citealt{Toonen2012}]{SeBa}.
The initial stellar population is obtained from a Monte Carlo based approach, assuming a binary fraction of 50\%, the Kroupa Initial mass function \citep[IMF,][]{KroupaIMF}, a flat mass-ratio distribution, a thermal eccentricity distribution \citep{Heggie1975} and a flat orbital separation distribution in logarithmic space \citep{Abt1983}.
Finally, for each binary we assign an inclination angle $i$, drawn from the uniform distribution in $\cos i$.
We summarize the distributions of the the initial binary parameters and their ranges in Table 1.

To take into account the star formation history of the Galaxy, we exploit a code originally developed by \cite{Nelemans2001a, Nelemans2004} and updated by \cite{Toonen2013}, in which the star formation rate is modeled as a function of time and position based on \citet[][see Equations (1) - (3) of \citealt{Nelemans2004}]{BoissierPrantzos1999}.
The absolute magnitudes for WDs are deduced from the WD cooling curves of pure hydrogen atmosphere models \citep[][and references therein\footnote{See also http://www.astro.umontreal.ca/~bergeron/
CoolingModels.}]{Holberg2006,Kowalski2006,Tremblay2011}.
To convert the absolute magnitudes to observed magnitudes (e.g. for the Sloan $r$ band) we use the following expression:
\begin{equation}
r_{\rm obs} = r_{\rm abs} + 10 + 5\log d + 0.84 A_{\rm V},
\end{equation}
where $d$ is the distance to the source in kpc, $0.84 A_{\rm V}$ is the extinction in the Sloan $r$ band, obtained from the extinction in the $V$ band, $A_{\rm V}$.
To compute the value of $A_{\rm V}$ at the source position, defined by the Galactic coordinates $(l,b)$ at the distance $d$, we use
\begin{equation}
A_{\rm V} (l, b, d) = A_{\rm V} (l,b) \tanh \left(\frac{d \sin b}{h_{\rm max}} \right), 
\end{equation}
where $A_{\rm V} (l,b)$ is the integrated extinction in the direction defined by $(l, b)$ 
from \citet{Schlegel1998}, $h_{\rm max} \equiv \min( h, 23.5 \times \sin b)$ and $h = 120$ pc is the Galactic scale
height of the dust \citep{Jonker2011}.
To convert $r$ magnitudes into Gaia $G$ magnitude we applied a colour-colour polynomial transformation with coefficients according to \citet[][table 6]{Carrasco2014}.
Finally, for our simulation we apply a magnitude limit of $r = 70$.
This limit is chosen to ensure that the simulated population can also be used for the GW detection simulations.

%%%%%%%%%%%%%%%%%%%%%%%%%%%%%%%%%
\begin{table}
\centering
\caption{Distribution of the initial binary parameters.}
\begin{threeparttable}
\begin{tabular}{|l|c|c|}
\hline 
Parameter           & Distribution & Range of definition                   \\ \hline \hline
Primary mass & Kroupa IMF\tnote{(a)}   & $0.95 < M$ M$_{\odot} < 10$\\ \hline
Binary mass ratio   & uniform in q      & $0< q \le 1$                          \\ \hline
Orbital separation  & uniform in  $\log a$\tnote{(b)}   & $0 \le \log \frac{a}{{\text R}_{\odot}}\le 6$ \\ \hline
Eccentricity        & thermal\tnote{(c)}       & $0 \le e \le 1$                      \\ \hline
Inclination         & uniform in $\cos i$     & $0 \le \cos i \le 1$                  \\ \hline
\end{tabular}
\begin{tablenotes}[para] 
\item[(a)]\citet{KroupaIMF};
\item[(b)]\citet{Abt1983};
\item[(c)]\citet{Heggie1975};
\end{tablenotes}
\end{threeparttable}
\end{table}
%%%%%%%%%%%%%%%%%%%%%%%%%%%%%%%%%%

There are at least two phases of mass transfer in the standard picture of formation of a DWD system. 
To form a short-period DWD binary at least one mass transfer phase needs to be a common envelope \citep{Paczynski1976,Webbink1984}.
In our simulation we adopt two evolutionary scenarios, with two different treatments of the CE phase: the $\alpha \alpha$ and the $\gamma \alpha$ scenarios.
In the $\alpha \alpha$ scenario the CE phase is described by the so-called $\alpha$-formalism \citep[see][for review]{Ivanova2013}.
In this prescription, the CE outcome is determined by the conservation of the orbital energy \citep{Webbink1984}, where $\alpha$ represents the efficiency in the exchange of the orbital energy and the binding energy of the envelope, described by another free parameter of the model $\lambda$.
The two parameters can be combined using equations (2) and (3) of \cite{Toonen2013} to a single unknown $\alpha \lambda$, that based on \citet{Nelemans2000} we adopt to be $\alpha \lambda = 2$.
In the second scenario, proposed in order to explain properties of observed DWDs, the CE is described by an alternative $\gamma$ parametrization \citep{Nelemans2000, Nelemans2005}.
In the $\gamma$-formalism the binary orbital evolution is driven by angular momentum loss, that is carried away through the mass loss process, and $\gamma$ is the efficiency of this mechanism.
For this  scenario we assume the value of the $\alpha\lambda$ as in the $\alpha\alpha$ CE model and $\gamma =1.75$ \citep{Nelemans2000}. 
In the $\gamma \alpha$ prescription $\gamma$-formalism is applied whenever a binary does not contain a compact object or when the CE is not driven by a tidal instability, in which case $\alpha$ prescription is used. 
Thus, in the $\gamma \alpha$ scenario, the first CE is typically described by the $\gamma$ formalism and the second by the $\alpha$ formalism.

The main differences between the two populations obtained with these different prescriptions are: the total number of binaries and their mass ratio distribution.
Using the $\gamma \alpha$ model one typically obtains twice as many binaries compared to the $\alpha \alpha$ scenario. 
Moreover, the mass ratio distribution in the $\gamma \alpha$ spans a wider range of values, which agrees better with the currently observed DWD population, while the majority of the population formed via $\alpha \alpha$ scenario will show mass ratio around 0.5 \citep[see][figure 2]{Toonen2012}.
This is due to the fact that in the $\alpha$ prescription the orbit always shrinks significantly. 
While when using the $\gamma$ prescription the CE outcome depends strongly on the binary mass ratio \citep[see, e.g., Equation (A.16) of][]{Nelemans2001a}:
for a roughly equal mass binary the orbit does not change much,
however, for a binary with a very different mass components the orbit shrinks strongly.

%%%%%%%%%%%%%%%%%%%%%%%%%%%%%%%%%%%%%%%%%%%%%%%%%%%%%%%%%
%%%%%%%%%%%%%%%%%%%%%%%%%%%%%%%%%%%%%%%%%%%%%%%%%%%%%%%%%

\section{EM detection} 

In this section we focus our analysis on two instruments: Gaia and the LSST.
Being photometric variability surveys, both are  expected to mostly detect new DWDs through eclipses \citep{Eyer2012}, and thus selecting mainly short period ones.
These DWDs are the most interesting for studying the final stages of binary evolution.

Gaia is a space mission, launched on the 19 December 2013, whose primary goal is  to provide a detailed 3D distribution and space motion of a billion stars in our Galaxy \citep{Gaia}.
During 5 years of mission Gaia will deliver positions, parallaxes, and proper motions for all stars down to $G \simeq 20$ over the whole sky.
According to the GUMS (Gaia Universe Model Snapshot) simulation Gaia will see between 250 000 and 500 000 WDs, and more than $60 \%$ of them will be in binaries \citep{Carrasco2014}.
Astrometrical and multi-colour photometrical observations will be possible for the Galactic WD population, but to fully characterize these sources  ground-based spectroscopic follow-up will be necessary\citep{Carrasco2014, Gaensicke2015}.
The majority of the Galactic WD population is too faint for the Radial Velocity Spectrometer (RVS) on board of the Gaia satellite, and even the brightest ones ($G < 15$) are typically featureless in RVS wavelength range.
Thus no radial velocities will be available for these sources.

The LSST is a ground-based telescope, currently under construction and expected to be fully operational in 2022 \citep{LSST}.
It will complement the Gaia study of the Milky Way stellar population down to $r \simeq 24$, with a possibility to extend this photometric limit down to $r \simeq 27$ with image stacking techniques.
The LSST will detect about 10 billion stars up to distances of $\sim 100$ kpc over half of the sky.
In particular, it will allow the discovery of several millions of WDs \citep[][Chapter 6]{LSST}.

The technical characteristics of the two instruments used for our study (sky coverage, average cadence, limiting magnitude and visibility constraints of the survey, etc.)  are summarized in Table 2. 

%%%%%%%%%%%%%%%%%%%%%%%%%%%%%%
\begin{table}
\begin{centering}
\caption{Gaia and the LSST technical characteristics. The quoted parameters are from \citet{Gaia} and \citet{LSST}.}
\label{tab:1}
\centering
 \begin{tabular}{ l | c | r }
 \hline
     & Gaia & LSST \\ \hline \hline
   Sky coverage & whole sky & $\sim 1/2$ sky  \\ \hline
   Wavelength coverage & 330-1050 nm & $ugrizy$ \\ \hline
   Bright limit & - & $r \simeq 16-17$ \\ \hline
   Depth per observation & $G \simeq 20.7 $ & $r \simeq 24$ \\ \hline
   Syst. photometric error (mag) & $0.001$ & $0.005$  \\ \hline
   Integration time (sec)& 40.5 & 15 + 15  \\ \hline 
   Nominal mission lifetime & 5 yr & 10 yr \\ \hline
   Average number of observations  & 70 & $10^3$  \\ \hline
   Average cadence of observations & 1 in 26 days & 1 in 3 days  \\ \hline
  \end{tabular}
\end{centering}
\end{table}
%%%%%%%%%%%%%%%%%%%%%%%%%%%%%

%%%%%%%%%%%%%%%%%%%%%%%%%%%%%
\begin{figure*}
        \centering
	 \includegraphics[width=0.7\textwidth]{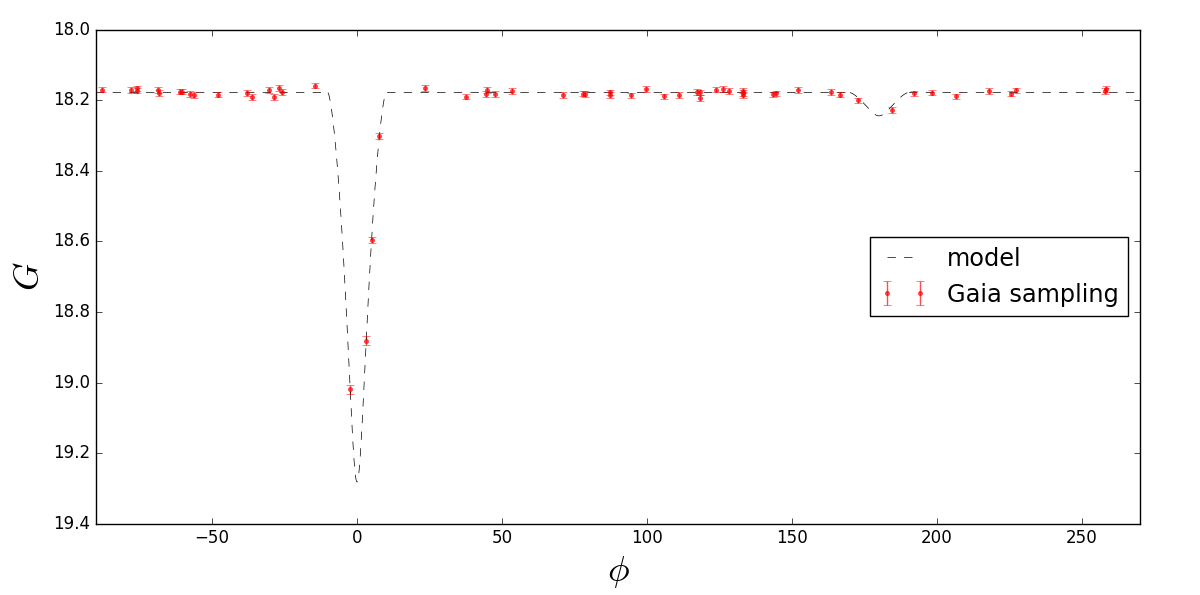} 
	 \includegraphics[width=0.7\textwidth]{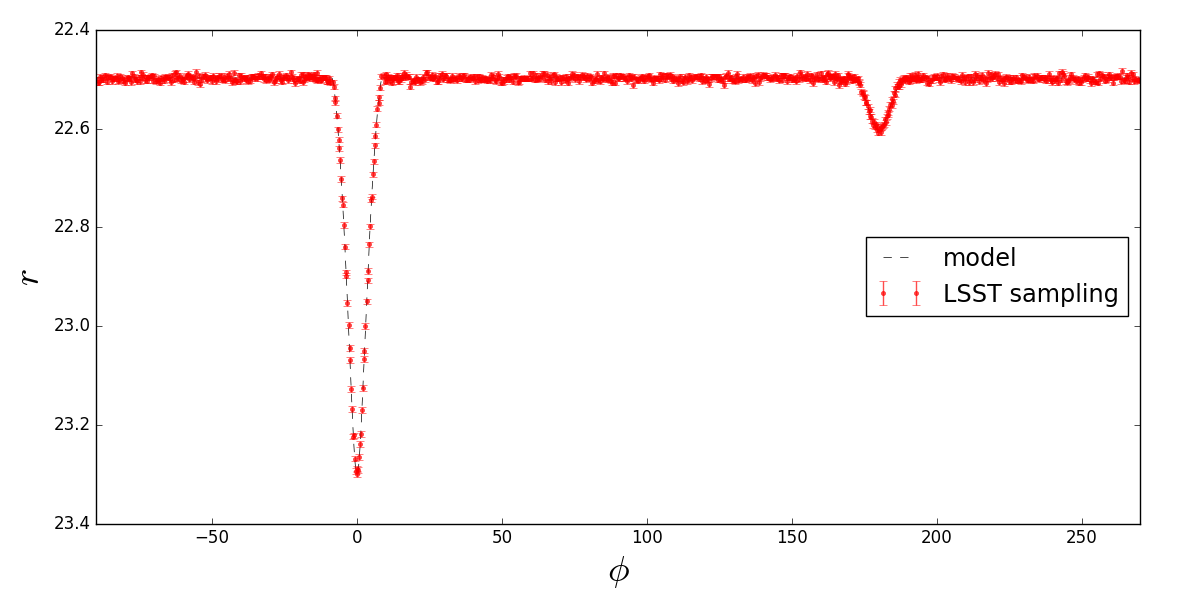}
         \caption{An example of phase folded light curves sampled with Gaia (top panel) and LSST (bottom panel) observations. The periods of the two sources are $P \simeq 21$ min and $P \simeq 24$ min respectively.}
       \label{fig:1}
\end{figure*}
%%%%%%%%%%%%%%%%%%%%%%%%%%%%%

%%%%%%%%%%%%%%%%%%%%%%%%%%%%%%%%%%%%%%%%%%%%%%%%%%%%%%%%%%%

\subsection{Simulations of light curves}

Next we simulate the light curves of the obtained DWD model population by using a  geometrical model, in which we compute the flux of a binary for given binary parameters: $a, R_1, R_2, r_1, r_2$ and $d$, where $a$ is the binary orbital separation, $R_1$ and $R_2$ are the respective radii of the two binary components, and $r_1$ and $r_2$ are their $r$-band magnitude. 
Note that in this work we adopt the definition of the primary as the brightest WD, and secondary as the dimmest WD of the pair.

In this simple treatment the limb darkening effect is neglected, so stars are considered spherically symmetric with a uniform surface brightness distribution.
Gravitational distortion (ellipsoidal variation) and mutual heating are also not taken into account.
Neglecting these effects implies looking for photometric variability caused by eclipses alone, that limits our search to systems with a very narrow range of inclination angles $i \sim 90^{\circ}$.
For DWDs the variation in the light curve induced by mutual heating is not expected to be significant, given the small size of WD stars and roughly equal size binary components. 
We estimate the maximum flux variation due to the mutual heating to be at most of the same order of magnitude as the average eclipse depth, if we assume the maximum efficiency for this process.
To test whether including the ellipsoidal variation in our simulation could enlarge the sample of detectable sources,
we estimate how many systems in our simulated population would show the maximum amplitude of the ellipsoidal variation greater than 1\% using the theoretical prediction from \citet{Hermes2012}:
\begin{equation}
\frac{L(\phi)}{L} = \frac{-3(15+u_1)(1+\tau_1)(R_1/a)^3(m_2/m_1)\sin^2 i}{20(3-u_1)} \cos(2\phi),
\end{equation}
where $L$ is the total luminosity of the system, $u_1= 0.1 - 0.5$ and $\tau_1=1.0$ are the limb-darkening and gravity-darkening coefficients for the primary, and $\cos(2\phi)=1$.
We find $\sim20$ systems (in both formation scenarios) with the maximum amplitude of ellipsoidal variation greater than 1\% in our simulation formed via $\alpha \alpha$ ($\gamma \alpha$) scenario with $G/r < 24$.
Thus, including ellipsoidal variation in our simulation would increase the number of detected system by at most a couple of tens new systems.

To evaluate the relative photometric error per single observation with Gaia we use:
\begin{equation}
\sigma_{\rm G} = 1.2 \times 10^{-3} (0.04895z^2+1.8633z+0.00001985)^{1/2},
\end{equation}
where $z=\max [10^{0.4(12-15)} , 10^{0.4(G-15)}]$ \citep[][Section 8.2]{Gaia}.
To evaluate an expected photometric error per single observation with the LSST we use 
\begin{equation}
\sigma_{\rm r} = (\sigma_{\rm sys}^2 + \sigma_{\rm rand}^2)^{1/2},
\end{equation}
where, according to \citet[][Section 3.5]{LSST}, $\sigma_{\rm sys}=0.005$ is the systematic photometric error, $\sigma_{\rm rand}^2 = (0.04-{\tilde \gamma})x+{\tilde \gamma} x^2$, $x=10^{(m-m_5)}$ is the random photometric error, $m_5$ and ${\tilde \gamma}$ are the $5\sigma$ limiting magnitude for a given filter and the sky brightness in a given band respectively.
Finally, we add a Gaussian white noise to our synthetic light curves.

The motion of the Gaia satellite is quite complex and cannot be expressed by an analytical formula: it is given by a combination of rotation of the satellite on its own axis, precession of the spin axis itself, and the revolution around the Sun \citep{Eyer2005}. 
Therefore, to get a realistic light curve sampling with Gaia, we used The {\it Gaia Observation Forecast Tool}\footnote{http://gaia.esac.esa.int/gost/}, that provides a list of observing times (TCB) per target for a given period of observation and target position on the sky.
To get a set of Gaia pointings for each binary in our simulation we use the largest available time interval that spans from 2014-09-26T00:00:00 TCB to 2019-06-01T00:00:00 TCB ($\sim$ 5 yr mission lifetime).
To simulate the light curve sampling with the LSST we use the anticipated regular cadence of 3 days over a nominal ten-year life span of the mission.
In Fig. 1 we show a comparison of the light curve sampling by Gaia (top panel) and LSST (bottom panel) for two binaries with similar orbital periods (21 min and 24 min).

%%%%%%%%%%%%%%%%%%%%%%%%%%%%%%%%%%%
\begin{figure}
        \centering
	 \includegraphics[height=0.38\textwidth, width=0.49\textwidth]{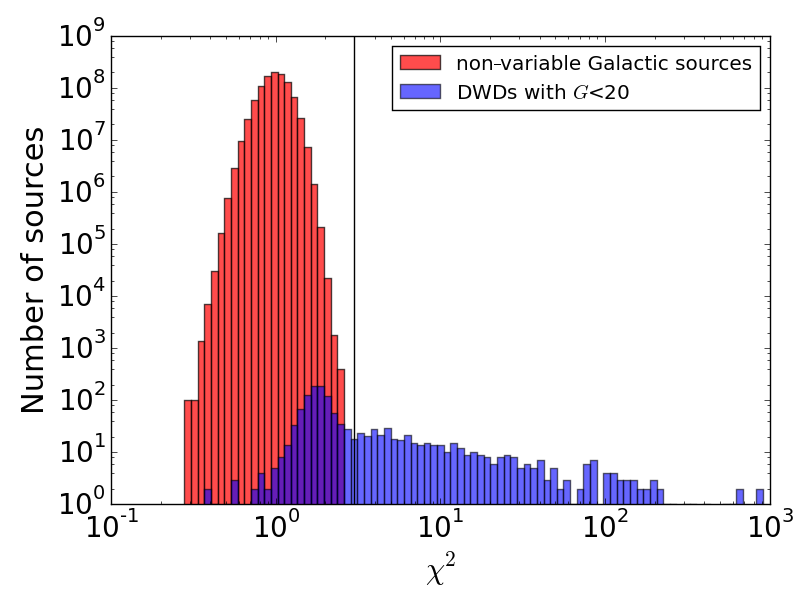}
         \caption{The $\chi^2$ distribution of simulated non-variable Galactic objects (in red) obtained using the classic apparent magnitude ($G$) distribution expected from star-counts (Prob. $\propto 10^{\bar{\gamma} G}$, where $0.2 \le \bar{\gamma} \le 0.4$), and the $\chi^2$ distribution of simulated DWDs in Gaia visibility range (in blue). The vertical line represents the threshold value $\chi^2 = 3$, above which we claim a detection.
}
       \label{fig:2}
\end{figure}

\begin{figure*}
        \centering
	 \includegraphics[ width=0.49\textwidth]{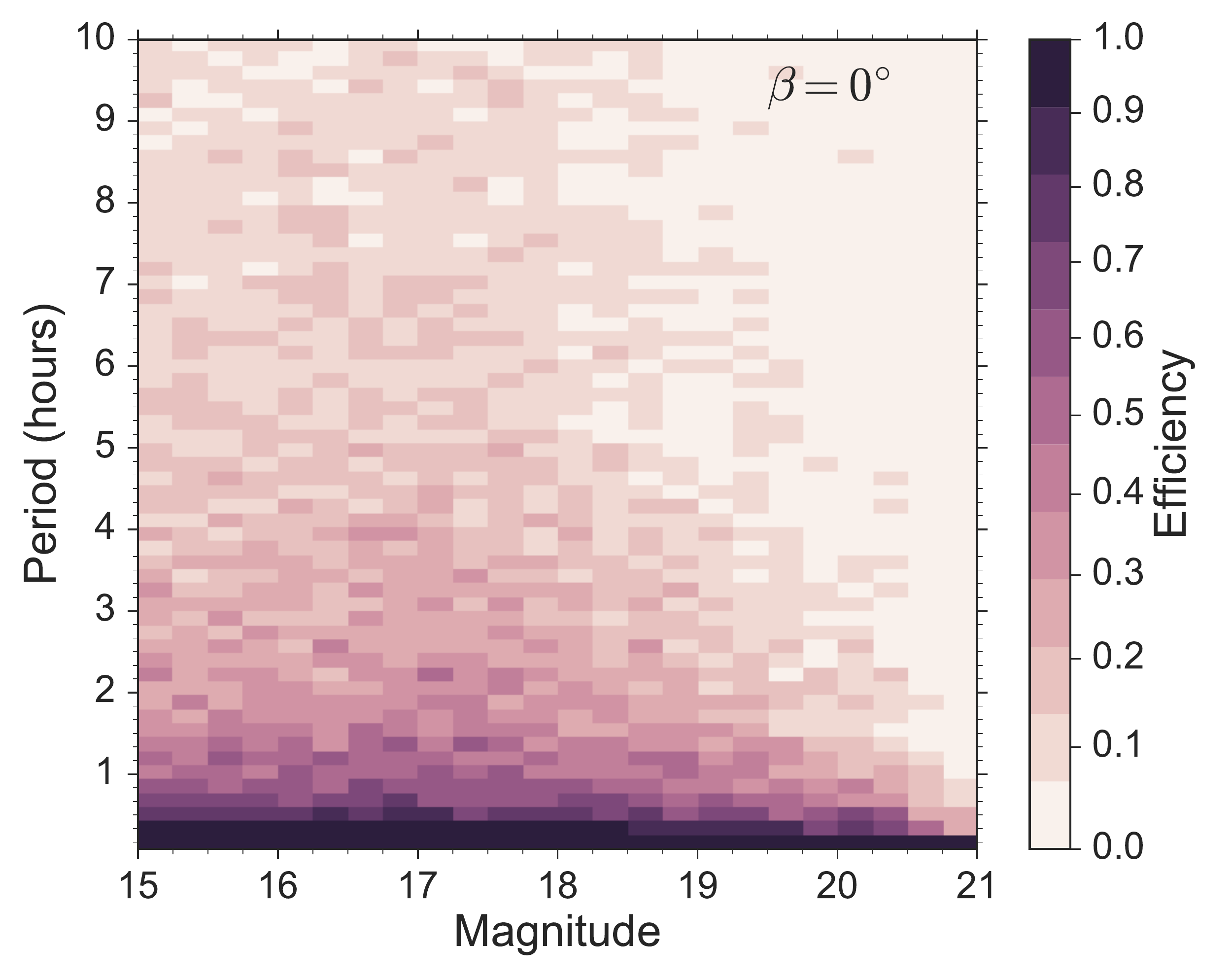}
	 \includegraphics[ width=0.49\textwidth]{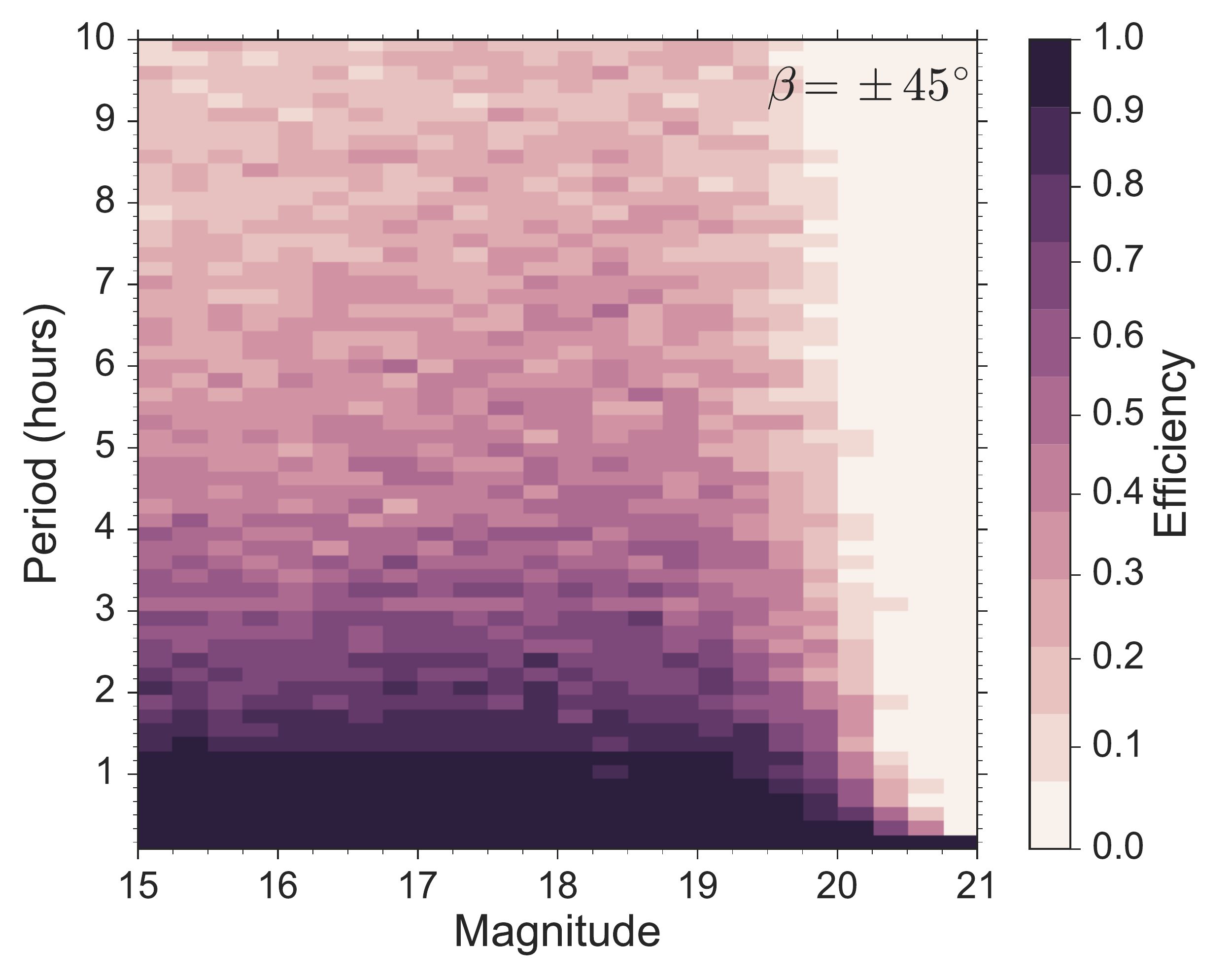}
	 \includegraphics[ width=0.49\textwidth]{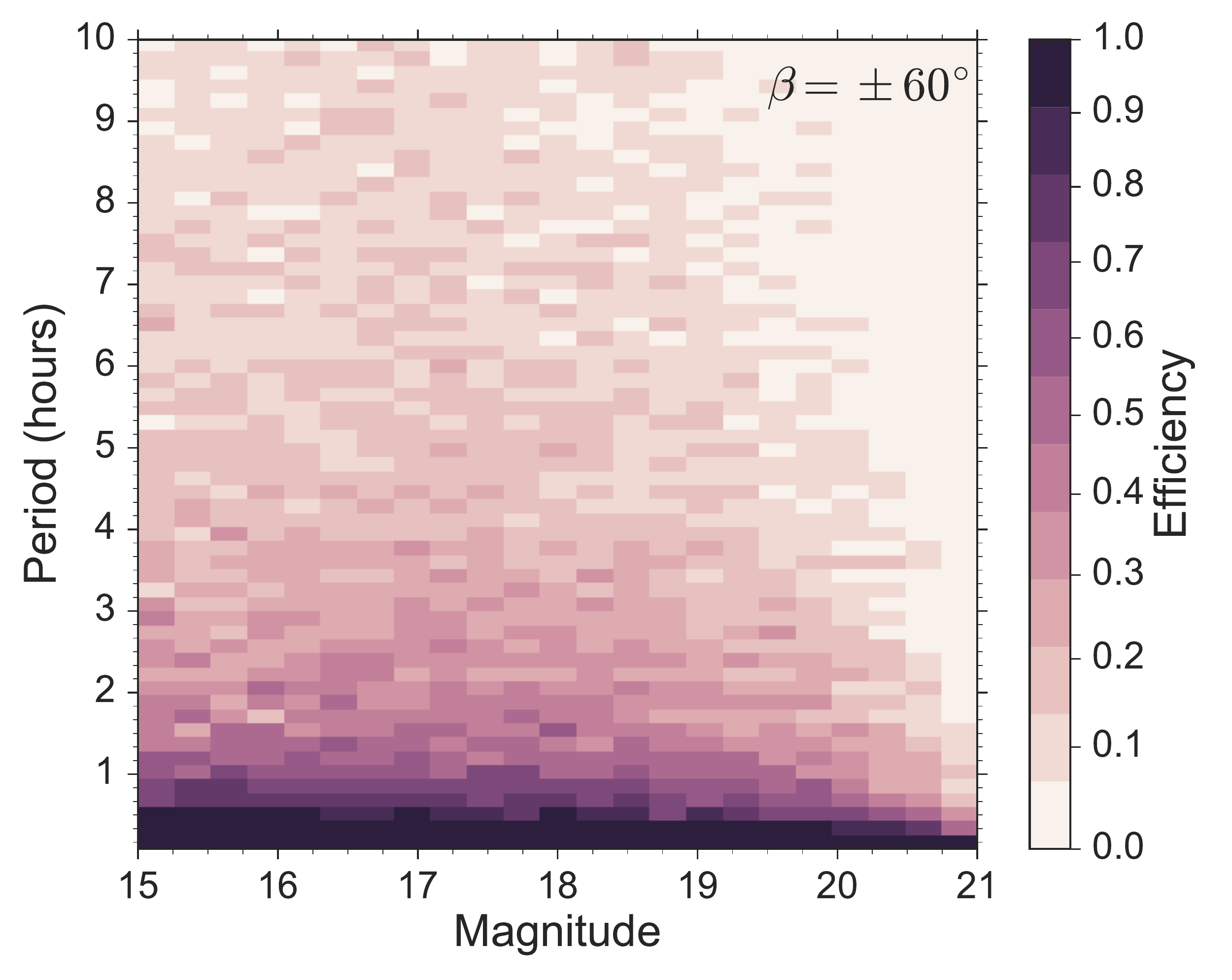}
	 \includegraphics[ width=0.49\textwidth]{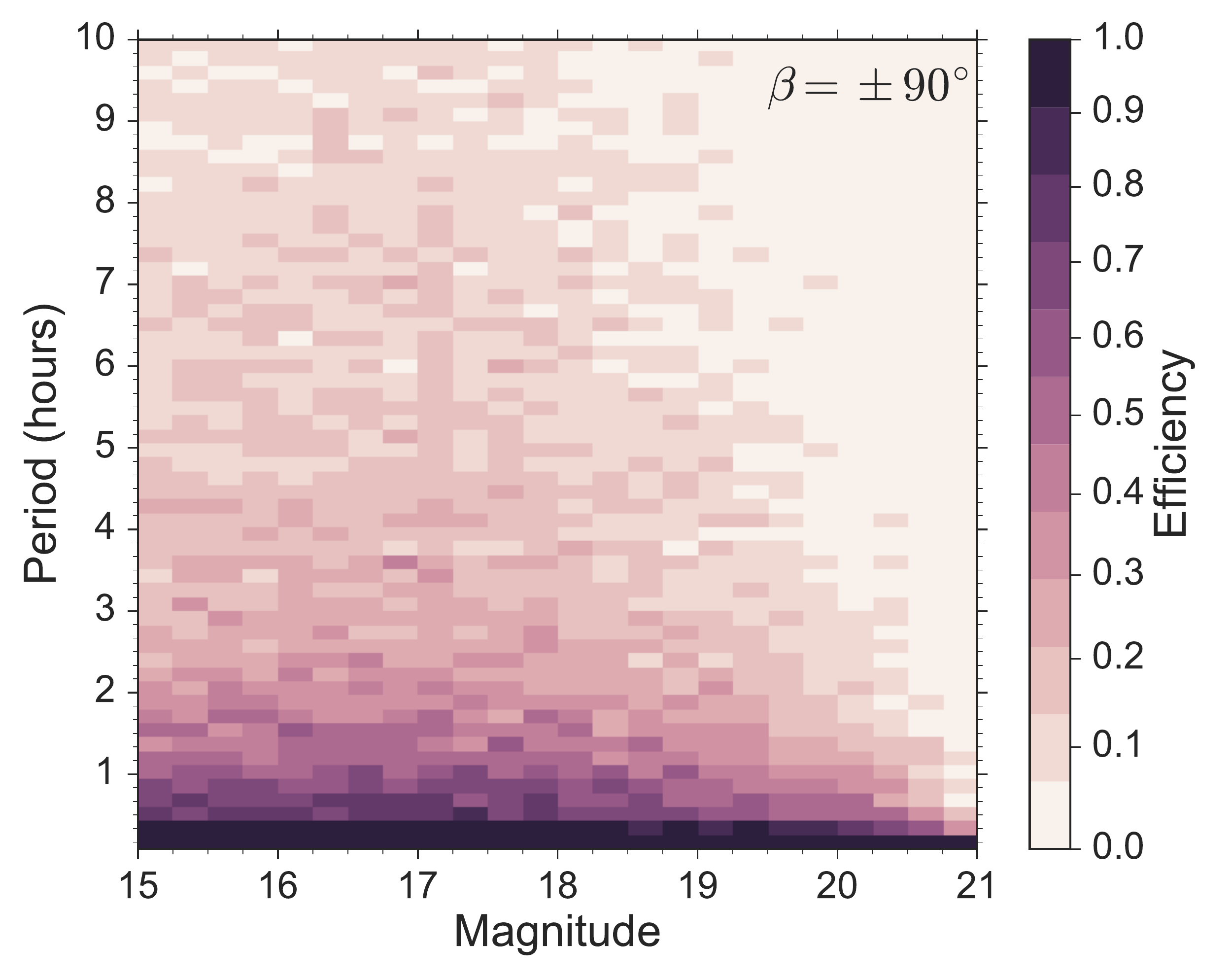}
         \caption{Detection efficiency of Gaia at $\beta = 0, +45, +60$ and $90^{\circ}$ ecliptic latitudes, that corresponds to respectively to 60, 200, 80 and 70 observations, computed for test binary system with $m_1 = 0.53$ M$\odot, m_2 = 0.35$ M$\odot, R_2 = 0.017$ R$\odot, R_1 = 0.8 R_2, d=1$ kpc  and $i = \pi/2$. The time step is 10 min and the magnitude step is 0.25. The colour indicates the instrument efficiency from 0 to 1. }
       \label{fig:3}
\end{figure*}

\begin{figure}
        \centering
	 \includegraphics[ width=0.49\textwidth]{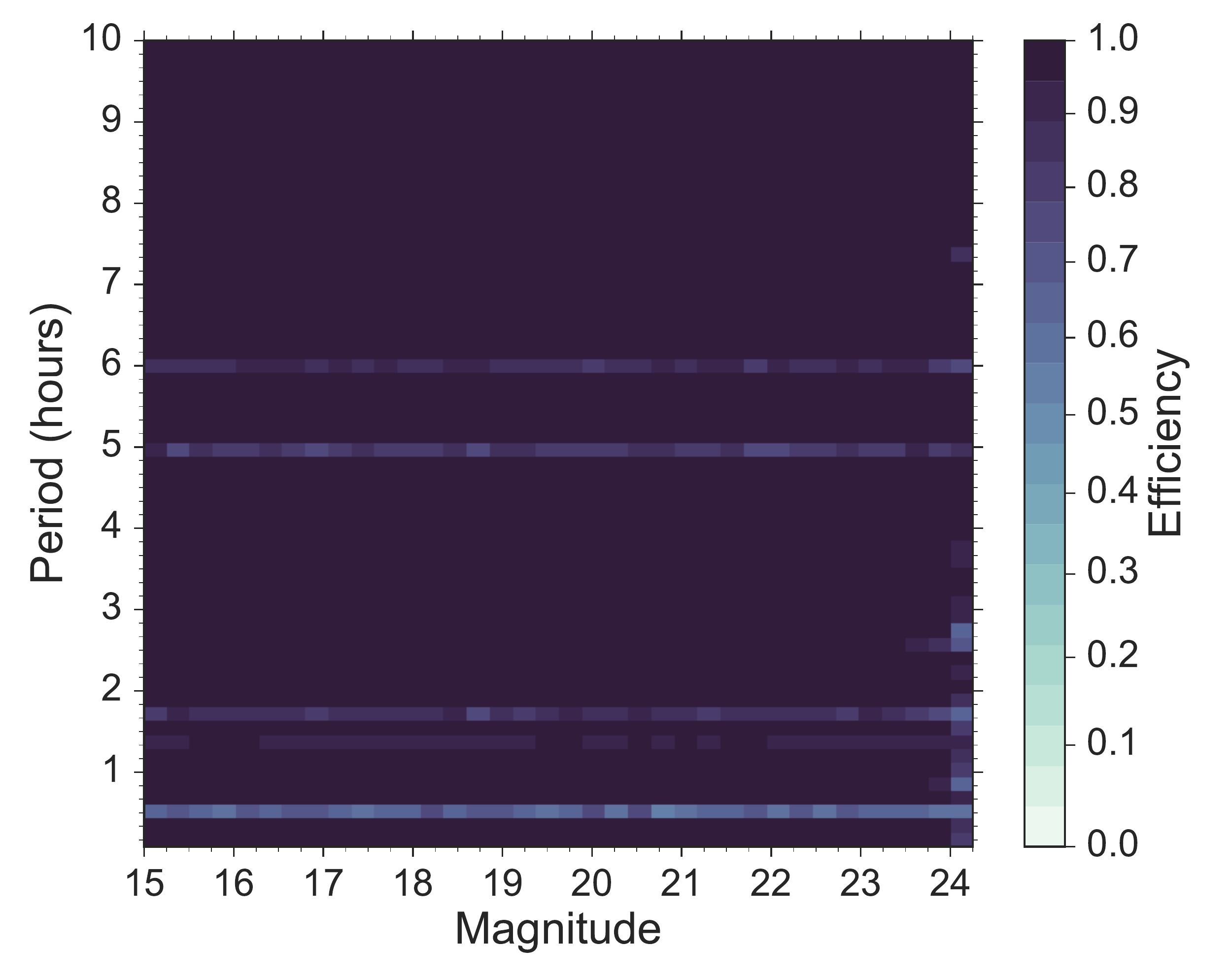}
         \caption{Detection efficiency of the LSST computed for test binary system with $m_1 = 0.53$ M$\odot, m_2 = 0.35$ M$\odot, R_2 = 0.017$ R$\odot, R_1 = 0.8 R_2, d=1$ kpc and $i = \pi/2$. The time step is 10 min and the magnitude step is 0.25. The colour indicates the instrument efficiency from 0 to 1. }
       \label{fig:4}
\end{figure}

\begin{figure}
        \centering
	 \includegraphics[width=0.43\textwidth]{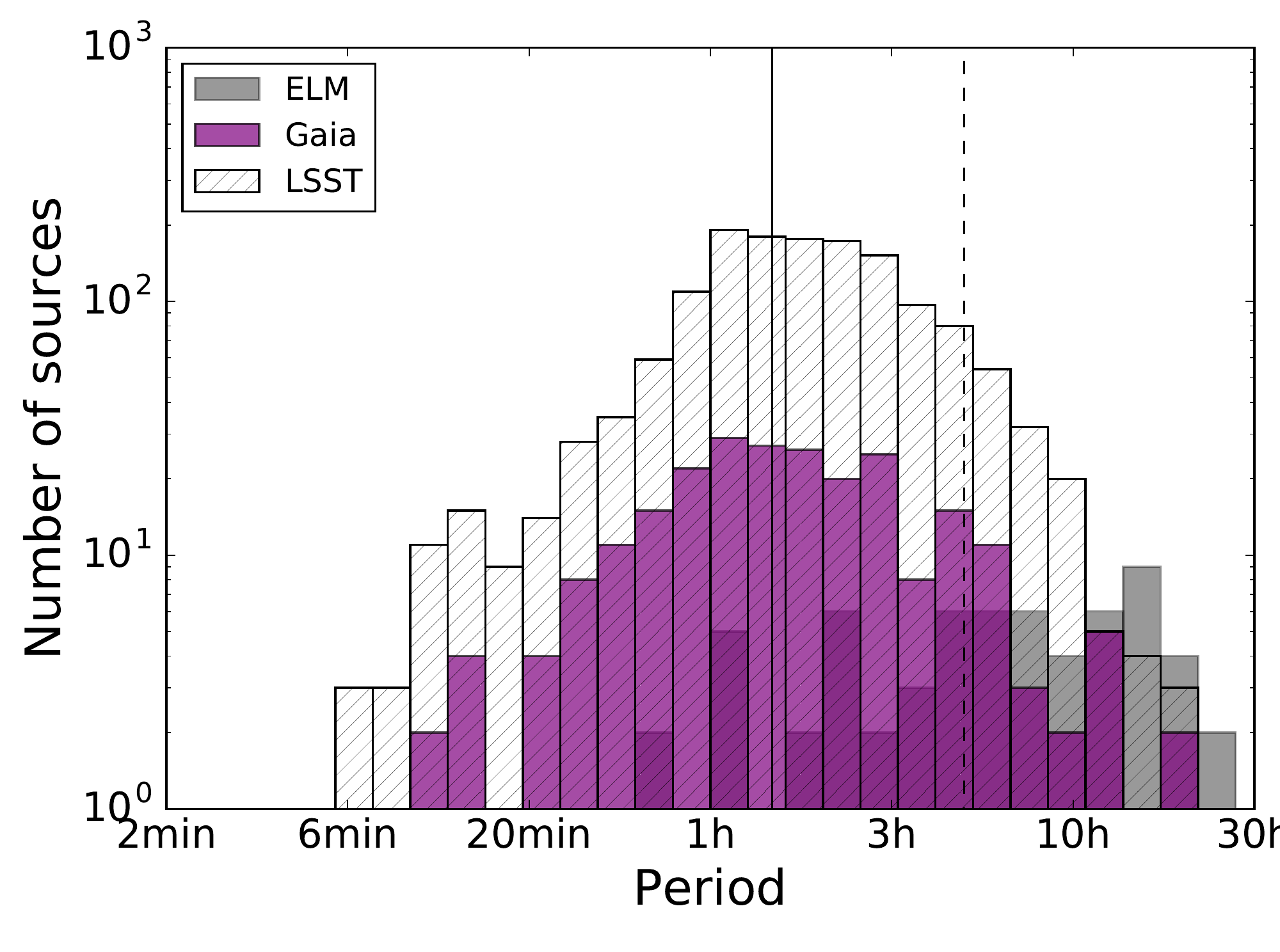}
         \caption{Number of detected sources as a function of the orbital period for the $\gamma \alpha$ formation scenario. The purple and hatched histograms represent respectively Gaia and LSST detections. The gray histogram shows binaries detected by the ELM survey taken from \citet{Gianninas2015}. The black continuous line represents the median of the detected periods in our simulation and dashed line marks the limit of the LISA band.}
       \label{fig:e}
\end{figure}
%%%%%%%%%%%%%%%%%%%%%%%%%%%%%%%%%%%

In order to count detections we applied the following criteria.
First, we check if the source presents variability by evaluating the $\chi^2$ value of the light curve with respect to the average source magnitude. 
To establish a $\chi^2$ threshold value above which we consider a source as variable, we compute the $\chi^2$ distribution of non-variable Galactic objects in the Gaia magnitude range.
The result is represented in Fig 2.
This simple test allows us to distinguish between variability due to a binary nature of the source and variability induced by photometric fluctuations of observations of non-variable objects.
In this simulation we did not take into account any other type of variable stars that will be present in the Galaxy such as pulsating WDs (DAVs: ZZ Ceti), Delta Scuti and SX Phoenicis stars, or variability due to deformation or heating in these binaries \citep[see for example][]{Macfarlane2015, Toma2016}.
In real data these stars will exhibit a similar behavior to eclipsing DWDs and will contaminate the sample of candidate DWDs.
Thus, in general additional analysis techniques will be required in order to confirm DWD candidates.
For the Gaia data this analysis will be done by the Gaia Data Processing and Analysis Consortium (DPAC) \citep{Eyer2014}.

It is evident from Fig. 2 that for $\chi^2 > 2$ there is little overlap between the population of non-variable sources (red histogram) and the population of eclipsing binaries. 
To be conservative we adopt  a threshold value of  $\chi^2 = 3$.
Finally, we require that a minimum number of data points,  $N_{\rm samp}$, with flux at least 3$\sigma$ below the out-of-eclipse level, falls within the eclipse phase: for Gaia we adopt $N_{\rm samp} \ge 3$ and for the LSST $N_{\rm samp} \ge 10$. 
This requirement introduces a constrain on the ratio between the duration of the eclipsing phase $t_{\rm ecl}$ and the binary orbital period $P$, such that $t_{\rm ecl}/P = N_{\rm samp}/N_{\rm tot}$, where $N_{\rm tot}$ is the total number of observations per source (see Table 2).
By using a geometrical argument $t_{\rm ecl}$ corresponds to the time it takes the occulting star to move twice the distance from the first contact (the point when the apparent stellar disks are externally tangent) to mid-eclipse (when stellar centres are aligned), so $t_{\rm ecl}/P$ can be find as
\begin{equation}
\frac{t_{\rm ecl}}{P} = \frac{\delta}{2\pi a},
\end{equation}
where $\delta = 2\sqrt{\left(R_1+ R_2 \right)^2 - a^2\cos^2 i}$ and $2\pi a$ is the total length of the orbit.
Note, that for an edge-on binary $\delta = 2 (R_1+R_2)$.
From eq.(6) we find that the typical $t_{\rm ecl}$ for a DWD binary in our simulated population is around 2 min.
Thus, we expect that Gaia will detect systems with typical periods $P \lesssim t_{\rm ecl} N_{\rm tot}/N_{\rm samp}\lesssim 45$ min. 
Following a similar reasoning one can anticipate that LSST will detect eclipsing binaries with $P \lesssim 3$ h.

%%%%%%%%%%%%%%%%%%%%%%%%%%%%%%%%%%%%%%%%%%%%%%%%%%%%%%%%%%%%

\subsection{Detection efficiency}

To assess the detection efficiency of the two instruments we simulate the sampling of a test light curve by varying the magnitude and period of a binary system with $m_1 = 0.53$M$_\odot, m_2 = 0.35$M$_\odot, R_2 = 0.017$R$_\odot, R_1 = 0.8 R_2, d=1$ kpc  and $i = \pi/2$.
The chosen parameters for the test light curve represent the average values of our simulated population.
For each period $P$ in the range between 5 min and 10 h (with 10 min steps) and magnitude ($r$ or $G$) between 15 and the photometric limit of the instrument (with 0.25 mag steps) we calculate 100 realizations of the test light curve sampling by randomly assigning the initial orbital phase.  
We determine whether the light curve was detected based on the criteria described in Sect. 3.1.
Finally, we represent the detection probability per bin as the number of times the test light curve was detected over 100 realizations.

As discussed in Sect. 3.1 our detection test depends on the total number of observations per source $N_{\rm tot}$.
For Gaia $N_{\rm tot}$ is uniform in ecliptic longitude $\lambda$ and has a strong dependence on ecliptic latitude $\beta$\footnote{https://www.cosmos.esa.int/web/gaia/table-2-with-ascii}: $N_{\rm tot}$ is minimum at $\beta \sim 0^{\circ}$, increases up to $\sim$ 200 observations per source at $\beta \pm 45^{\circ}$, and decreases down to $\sim 70$ at ecliptic poles $\beta \pm 90^{\circ}$ \citep{Eyer2005}. Gaia detection efficiency for $\beta = 0, +45, +60$ and $90^{\circ}$ ecliptic latitudes is represented in Fig. 3, where the impact of the different number  of observations is evident.
Figure 3 shows that for any fixed period (when the distance to the source is also fixed) Gaia generally detects more efficiently brighter binaries, simply because of the photometric performance of the instrument.
For example in the top left panel of Fig.3, for periods between 2-3 h one can see that the efficiency drops from 0.4 - 0.3 to 0 for increasing magnitudes.
However, for very short periods ($P \lesssim 20$ min) the efficiency remains approximately constant even at the faint end of the Gaia visibility range, independently of the number of observations.
At a fixed magnitude Gaia cadence works better for detection of short period sources: for $G=18$ the efficiency is > 0.4 for $P < 4$h and > 0.9 for $P < 30$ min (Fig.3 top left panel).
This is a consequence of the fact that the eclipse duration is fixed by the geometry of the system, so the time that the system spends in eclipse compared to the total orbital period is longer for systems with shorter periods (i.e., $t_{\rm ecl}/P$ decreases along the $y$-axis).
Thus, it is more likely to catch the binary in eclipse phase when the period of the binary is shorter.
By using this simple argument and  assuming a regular cadence of 70 observations one can preliminarily estimate the average number of detections by counting the number of DWDs in our synthetic population that satisfy $t_{\rm ecl}/P \ge 3/70$.
This gives around $250$ DWD systems with $G < 20.7$.

The efficiency of the LSST is illustrated in Fig. 4.
For the LSST we find that the average cadence of 1 observation in 3 days and the high number of data points make it very efficient at all magnitudes for all orbital periods $\lesssim$ 10 h (see Fig. 4).
Drops in efficiency visible in Fig. 4 (e.g. a horizontal stripe at 6 h) corresponds to periods that are submultiples of 72 h, the cadence of observations. 
As for Gaia, we estimate the number of binaries in our simulated population that can be positively detected with at least 10 observation per eclipse.
We find around $\sim 1.9 \times 10^3$ binaries with $r < 24$.

%%%%%%%%%%%%%%%%%%%%%%%%%%%%%%%%%%%%%%%%%%%%%%%%%%%%%%%%%%%%%%

\subsection{Results}

%%%%%%%%%%%%%%%%%%%%%%%%%%%%%%%%
\begin{figure*}
        \centering
	 \includegraphics[height=0.4\textwidth, width=0.49\textwidth]{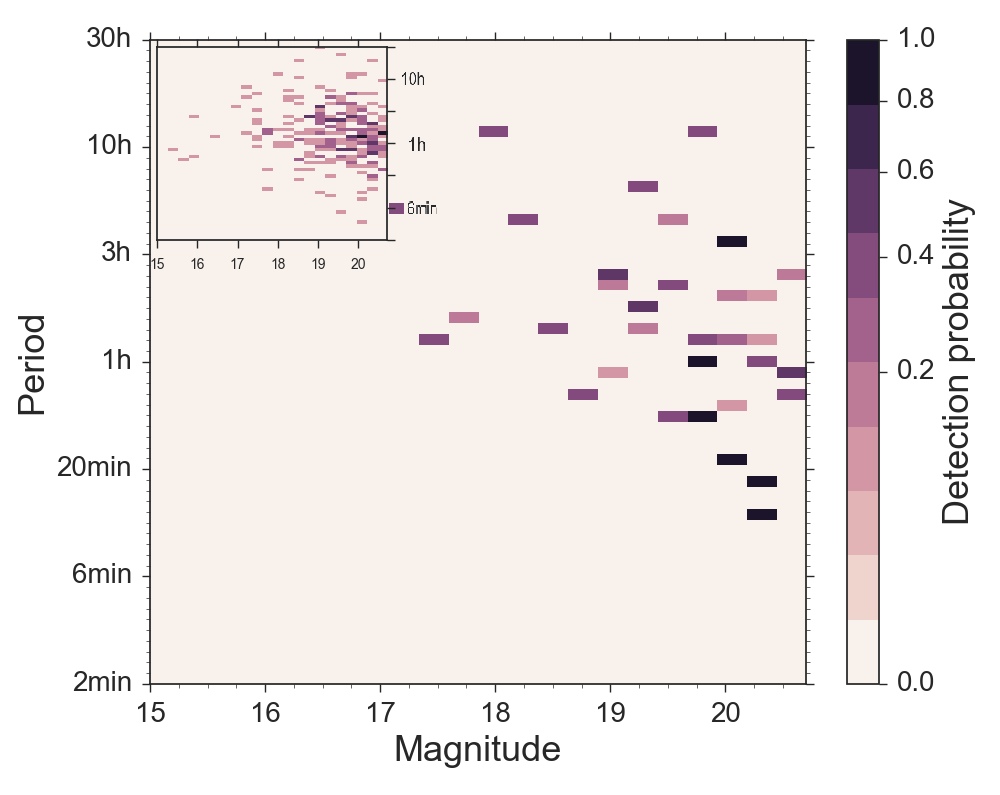}
	 \includegraphics[height=0.4\textwidth, width=0.49\textwidth]{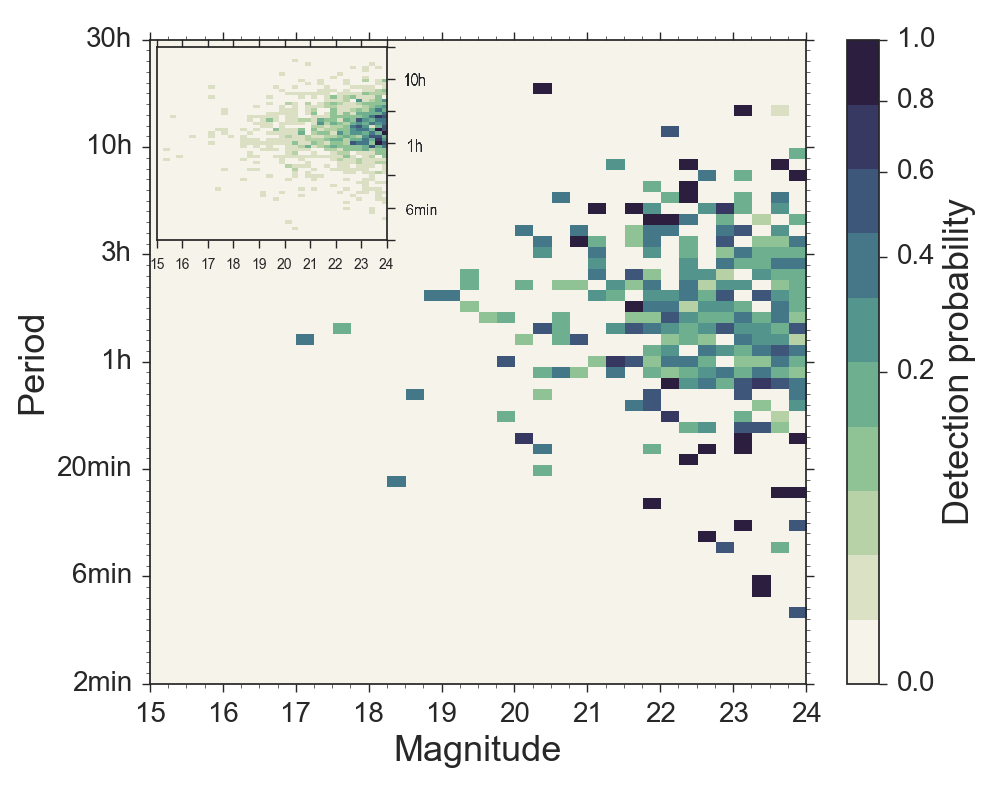}
	 \includegraphics[height=0.4\textwidth, width=0.49\textwidth]{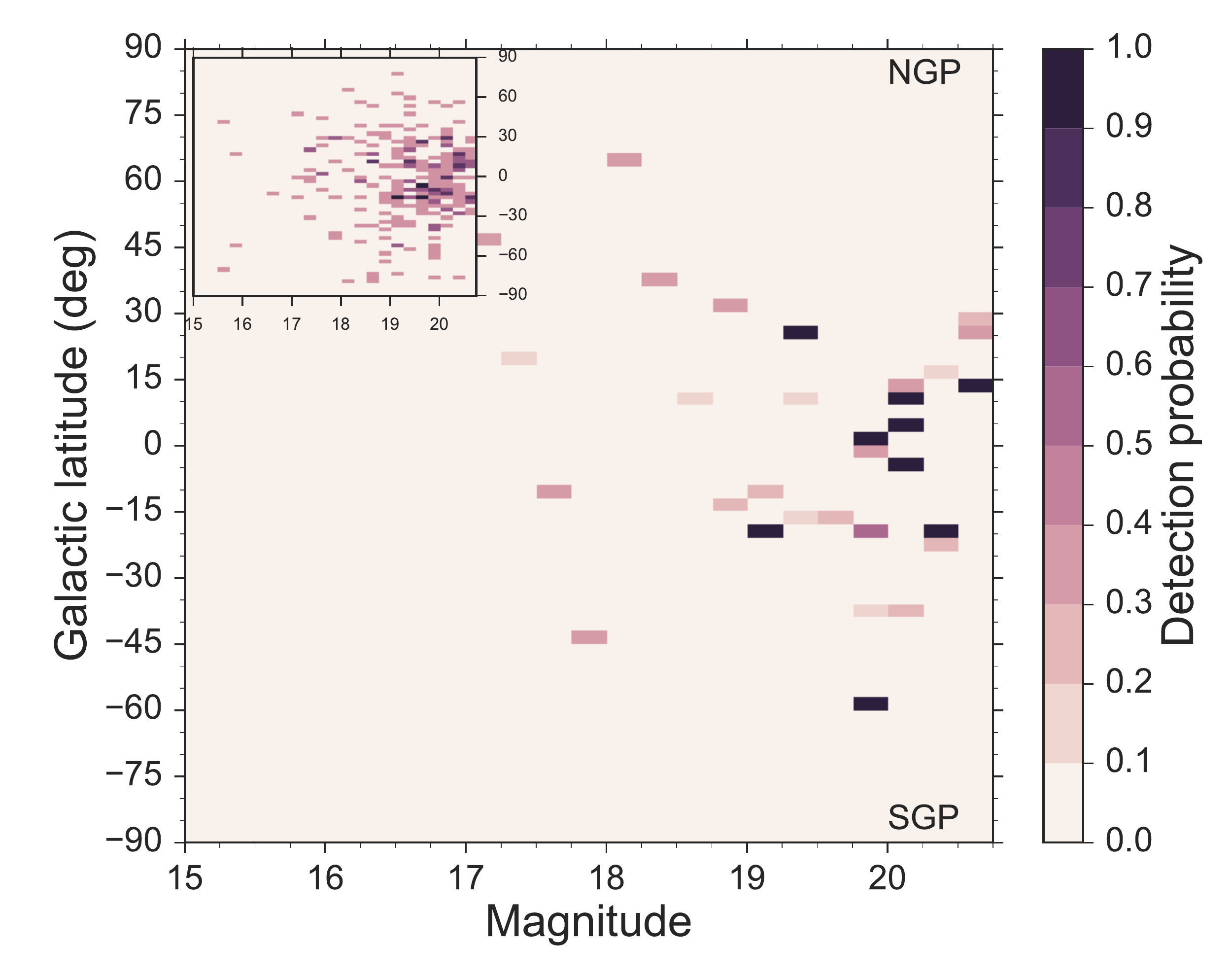}
	 \includegraphics[height=0.4\textwidth, width=0.49\textwidth]{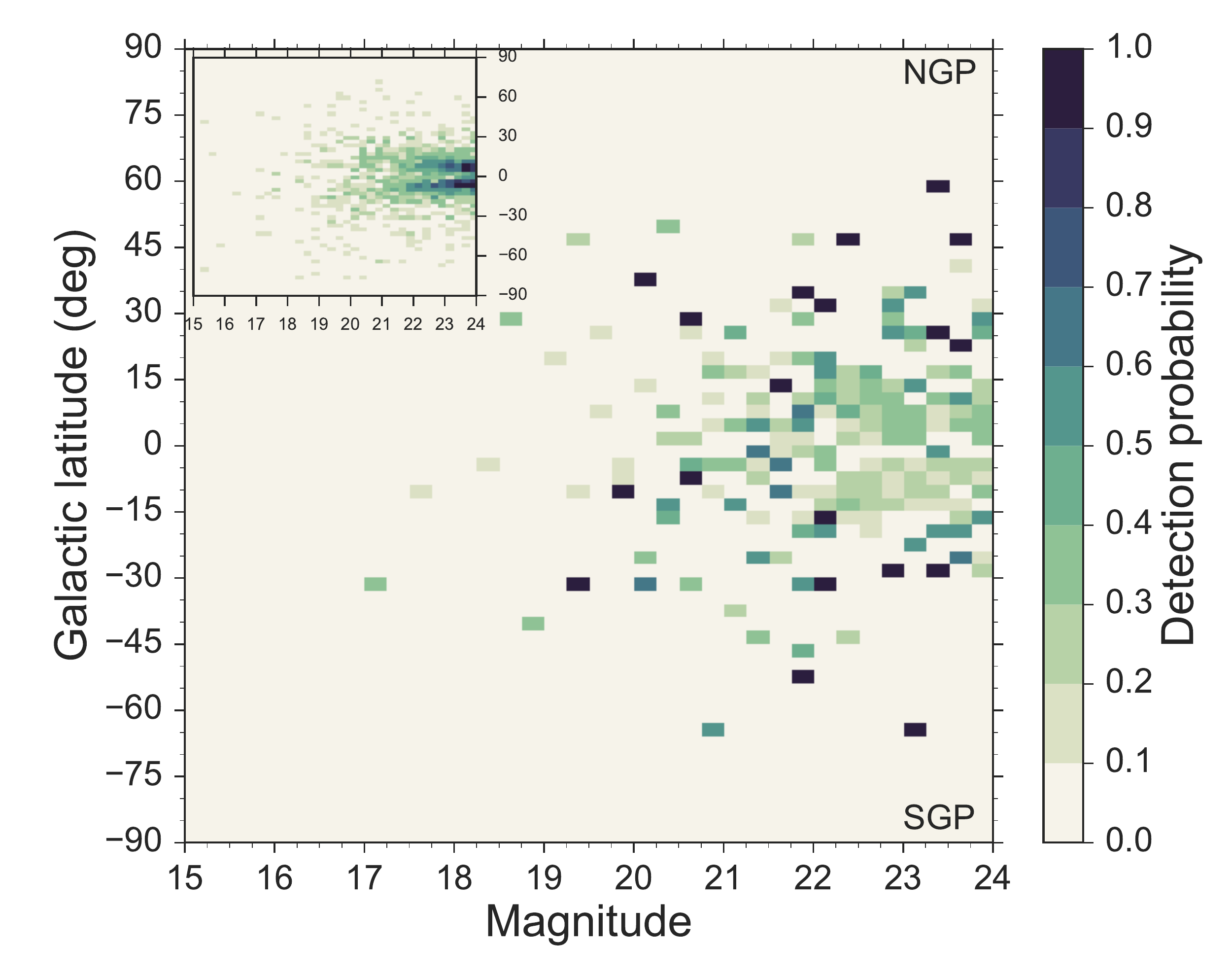}
	 \includegraphics[height=0.4\textwidth, width=0.49\textwidth]{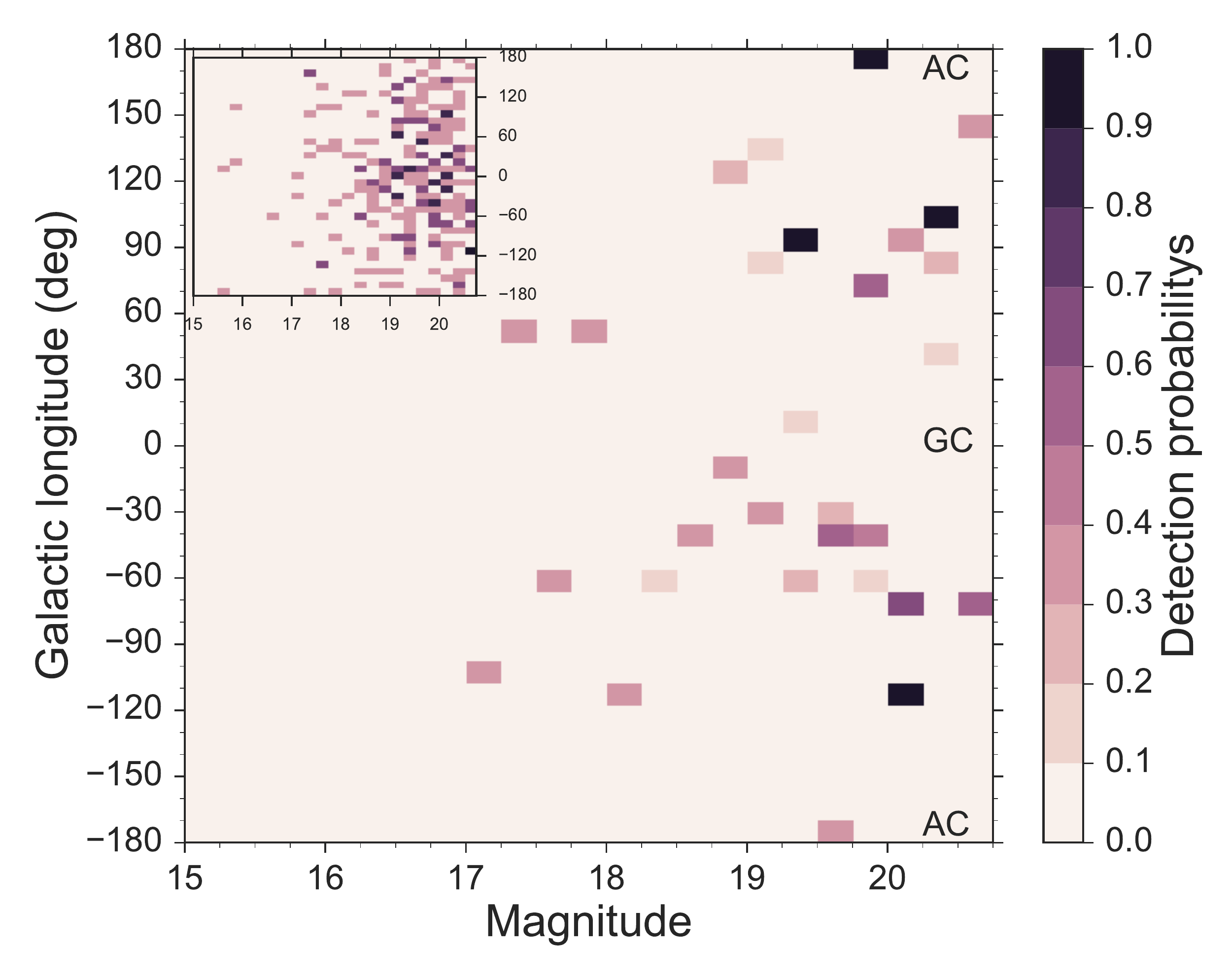}
	 \includegraphics[height=0.4\textwidth, width=0.49\textwidth]{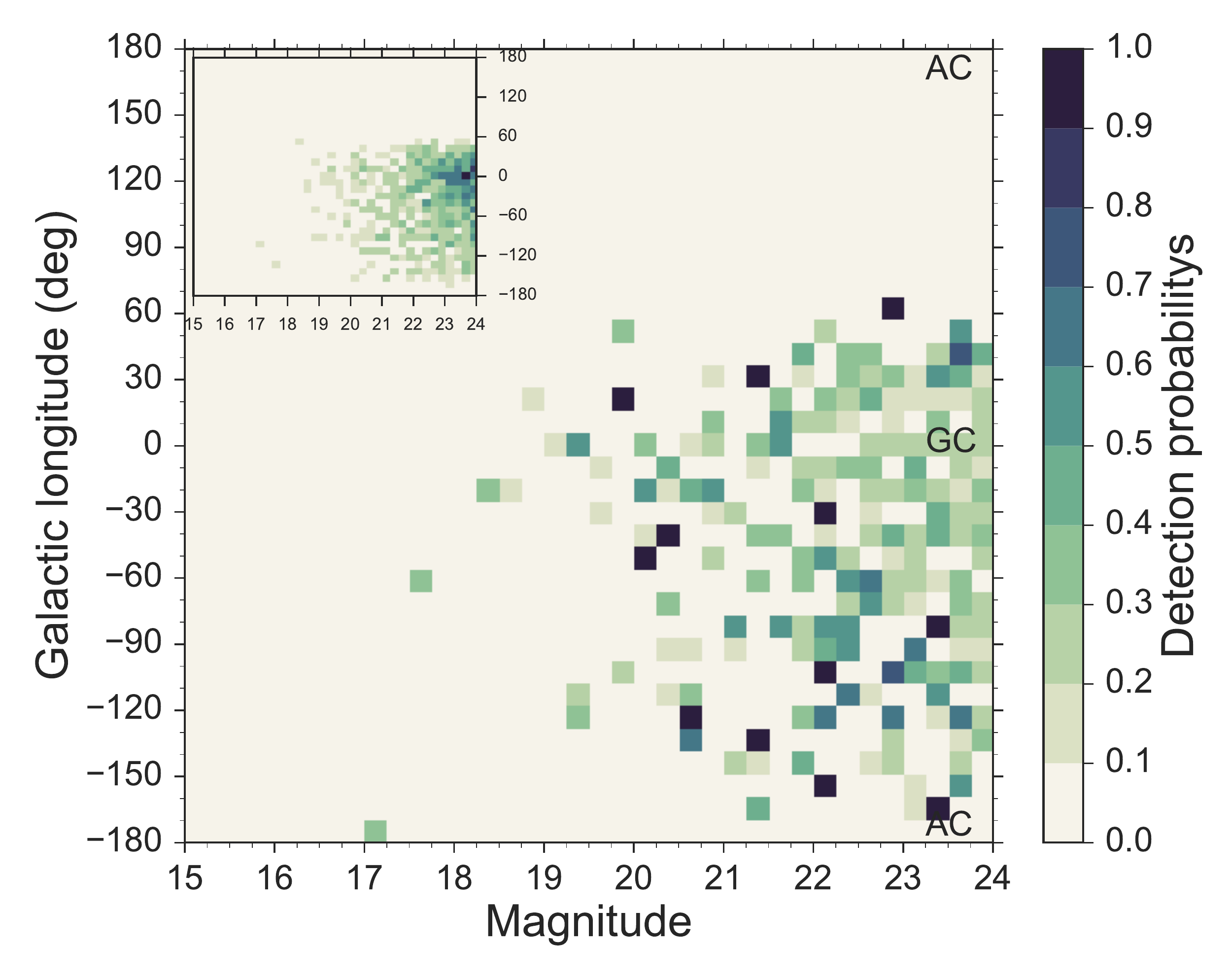}
         \caption{Contour plot for the number of detections in different 2D parameter spaces: left panels for Gaia, right panels for LSST. We show all the systems formed via the $\gamma \alpha$ scenario weighed by the probability of being detected. The respective inserts represent the distribution of all the systems with a non-zero detection probability. The colour indicates the detection probability: purple palette for Gaia and green palette for LSST. NGP ans SGP indicate the North and the South Galactic poles, GC and AC indicate Galactic centre and Galactic anti-centre. }
       \label{fig:5}
\end{figure*}
%%%%%%%%%%%%%%%%%%%%%%%%%%%%%%%%

For each binary in our simulated population we compute 100 light curve realizations by randomizing over the initial orbital phase, and 
we define the probability of detection as the fraction of times the light curve was positively detected (using the criterion described in Sect. 3.1) over the total number of realizations.
The following results pertain the fraction of the total Galactic DWD population that is: 1) above the photometric limit of the instrument,  2) for assigned orientation to the detector can be seen as eclipsing (i.e such that $\cos i \le (R_1+R_2)/a$), and 3) in a sky position covered by the survey.
In the reminder we call this population ``Gaia/LSST input population''.
Note, that the input population represents the maximum detectable sample for a given survey.

We find that 190 (250) binaries have a non-zero probability\footnote{A non-zero probability according to our definition means at least one detection in 100 (i.e. $\ge 0.01$).} to be detected by Gaia in the $\alpha \alpha$ ($\gamma \alpha$) scenario in 5 yr mission lifetime.
This represents $\sim$50\% of the Gaia input population in both formation scenarios.
Such detection percentage is due to a sparse Gaia sampling, that spread over the 5 yr mission time makes it difficult to detect systems with very narrow eclipses as in the case of DWDs (see Sect. 3.2).
The average number of detected binaries weighed by detection probability is 30 for the $\alpha \alpha$ and 50 for the $\gamma \alpha$ CE model respectively.
Essentially, Gaia will be sensitive to eclipsing binaries with orbital periods less than a few hours (50\% of these have periods < 1.6 h, see Fig. 5) up to the maximum of a few days.
The most distant binary detected is at the distance of 3.5 kpc.
In addition, we find that a possible extension of the Gaia mission up to 10 years \citep[][Section 5.3.2]{Gaia} will double the average number of detections compared to the nominal 5 yr mission lifetime.
Incidentally, we checked that our results are recovered even when we use a random sampling of the orbital phase interested of using detailed Gaia cadence.

Compared to Gaia, the ability of the LSST to see much fainter sources gives an order of magnitude more eclipsing binaries: 1100 (1460) DWDs have a non zero probability of being detected.
These detections represent $\sim$65\% (for both formation scenarios) of the LSST input population.
The average number of detected binaries weighed by the probability for the LSST is 850 (1167) DWDs for the $\alpha \alpha$ ($\gamma \alpha$) scenario.
The maximum distance in the LSST detected sample is $\sim$ 10 kpc.

Notably, half of the population detected by both instruments has periods shorter than  1.5 h as shown in Fig. 5. 
This  substantial subsample has orbital frequencies, $f = 1/P$, larger than 0.1 mHz, and thus is potentially detectable through GW radiation in the LISA band (see Sect. 4).
Both Gaia and LSST will enlarge the number of shorter period binaries, as the mean period of Gaia and LSST detections peaks around 1.5 h, while the mean period of the ELM binaries is 7.4 h \citep{Gianninas2015}.

In Fig. 6 we show the distribution of DWDs weighed by the probability of being detected by the instrument in different 2D parameter spaces: the magnitude-period distribution\footnote{Note, that we show only a part of the magnitude-period parameter space, where the majority of the detected population is located, while the whole range of detected periods extends up to a few days for both instruments, where the detections are sparsely distributed.} (top panels), magnitude-Galactic latitude (middle panels) and longitude (bottom panels) distributions, where colours trace the detection probability. 
The inserts in Fig. 6 represent the respective distributions of all sources with non-zero probability of detection.
Despite the fact that Gaia is more efficient at brighter magnitudes (Fig. 3), one can see that the majority of the detected population is faint ($G < 18$) and has periods less than a few hours ($P < 3$ h).
The former results reflects the magnitude distribution of the input population that peaks around the faint end of the Gaia visibility range, the later is a consequence of our detection criterion as discussed in Sect. 3.1.
Comparing the two upper panels in Fig. 6 it is evident that the LSST with its deeper photometric limit, has access to a much larger fraction of the total population.
In particular, while Gaia operates in the same magnitude range of the ELM ground-based optical survey, the LSST will extend the sample of known DWDs to lower magnitudes.
However, the follow-up spectroscopy of such a faint sources will be a challenge even for up-coming facilities.

In the middle and lower panels of Fig. 6, we represent the spatial distribution of Gaia and LSST detections in the Galaxy.
Because of its photometric limit, Gaia will see only the closest sources ($d_{\rm max} = 3.5$ kpc $\ll$ radius of Galactic disc), therefore the distribution in longitude is featureless. 
On the other hand, one can start to see the distribution of DWD around the Galactic plane (insert middle left panel), since 3.5 kpc is larger than the vertical extension of the Galactic disc.
The distribution of DWDs in the Galaxy will become potentially visible with LSST.
The concentration of detected binaries towards the Galactic plane represent the Bulge of the Galaxy with its characteristic gap around $0^{\circ}$ Galactic latitude due to extinction in the disc (insert middle right panel).
The location of the LSST in the southern hemisphere is reflected in the lack of sources for Galactic longitudes greater than 60 $^{\circ}$ in Fig. 6 (bottom right panel).

%%%%%%%%%%%%%%%%%%%%%%%%%%%%%%%%%%%
\begin{figure}
        \centering
	 \includegraphics[width=0.39\textwidth]{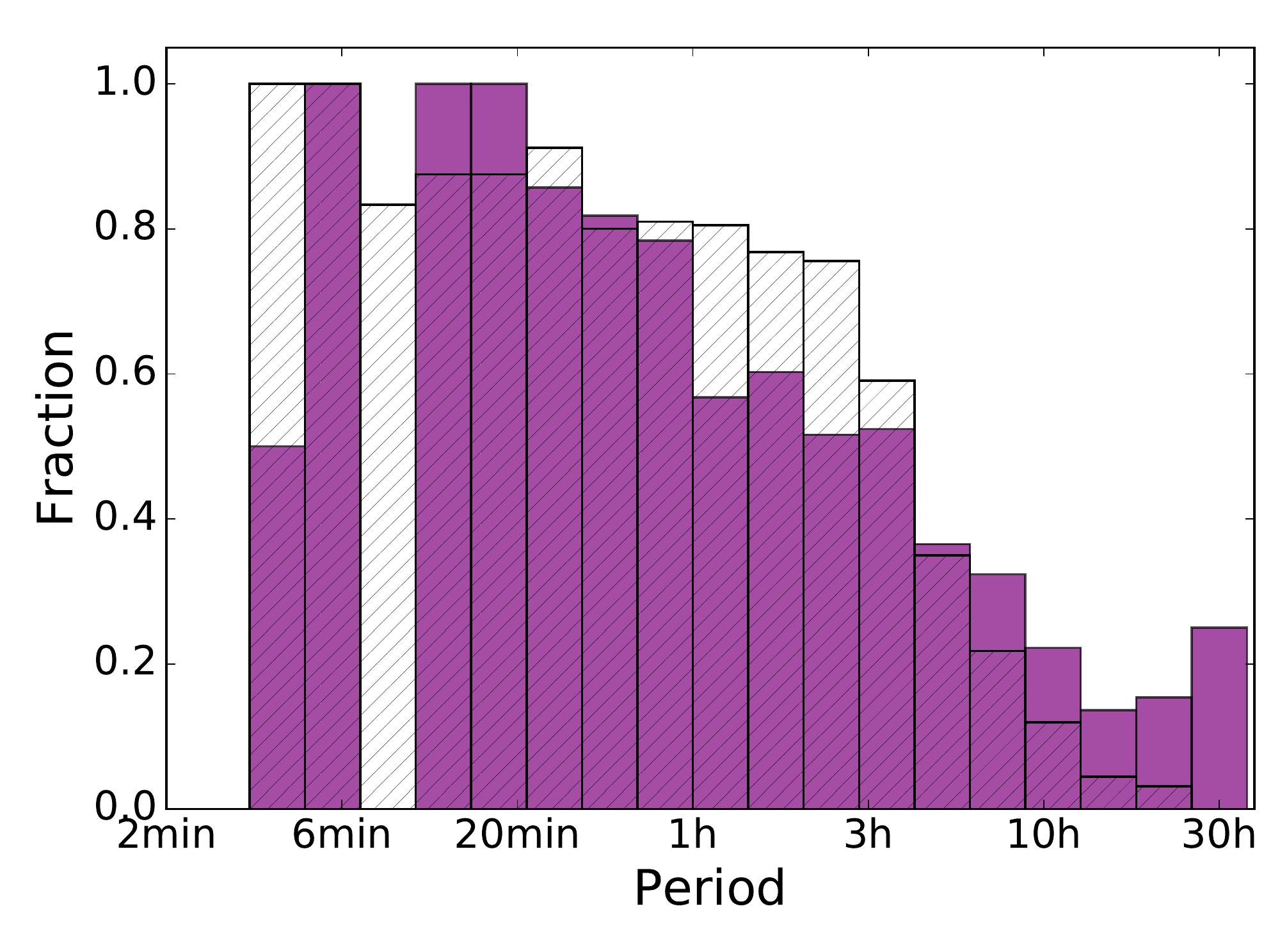}
	 \includegraphics[width=0.39\textwidth]{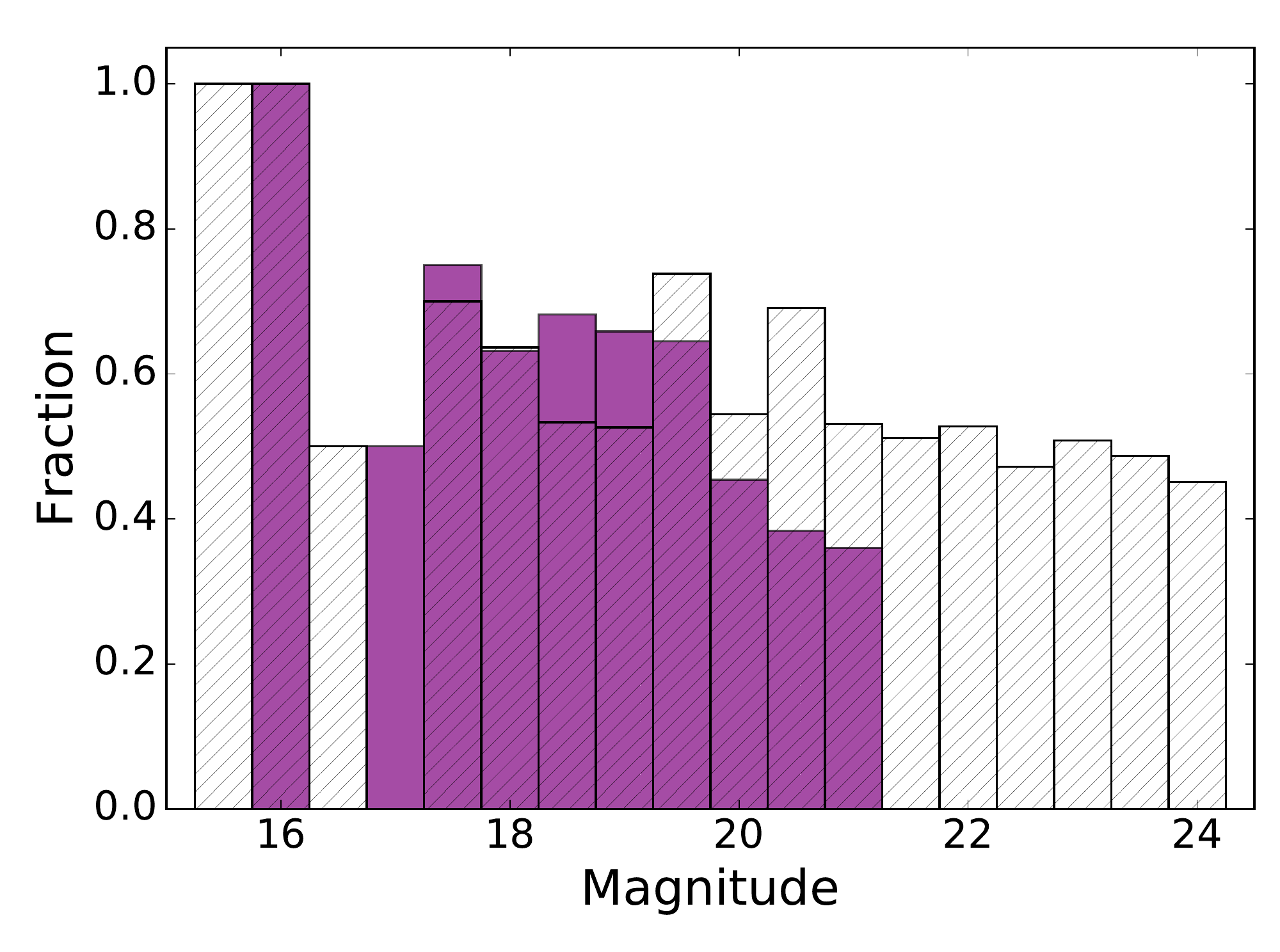}
	 \includegraphics[width=0.39\textwidth]{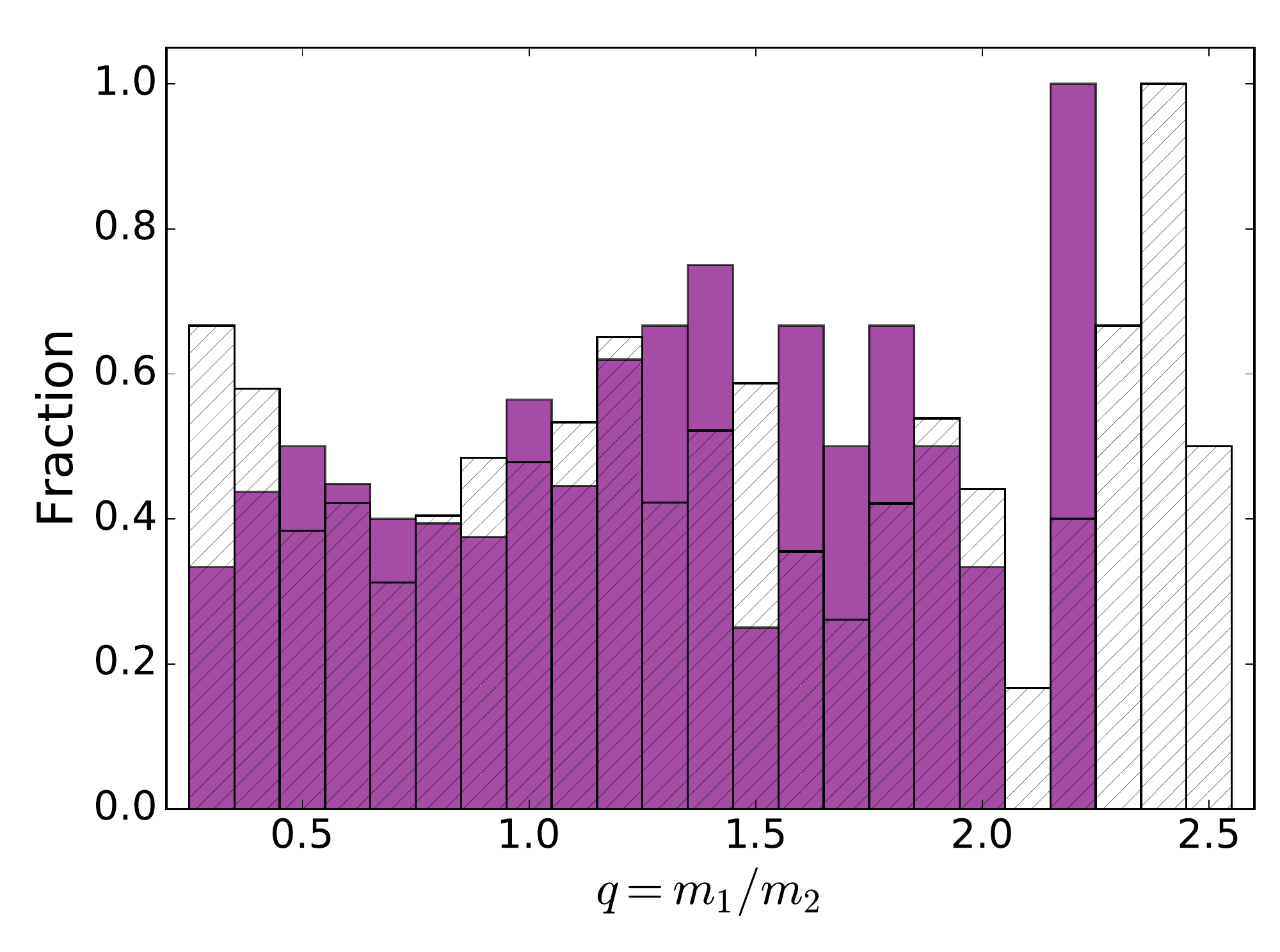}
         \caption{Number ratio of detected sources over the total input binaries per bin for the $\gamma \alpha$ formation scenario. From top to bottom we show the detection fraction as a function of period, magnitude and mass fraction $q$. The purple histogram shows Gaia detections and hatched histogram represents LSST detections.}
       \label{fig:6}
\end{figure}
%%%%%%%%%%%%%%%%%%%%%%%%%%%%%%%%%%%

%%%%%%%%%%%%%%%%%%%%%%%%%%%%%%%%%%%
\begin{figure}
        \centering
	 \includegraphics[width=0.4\textwidth]{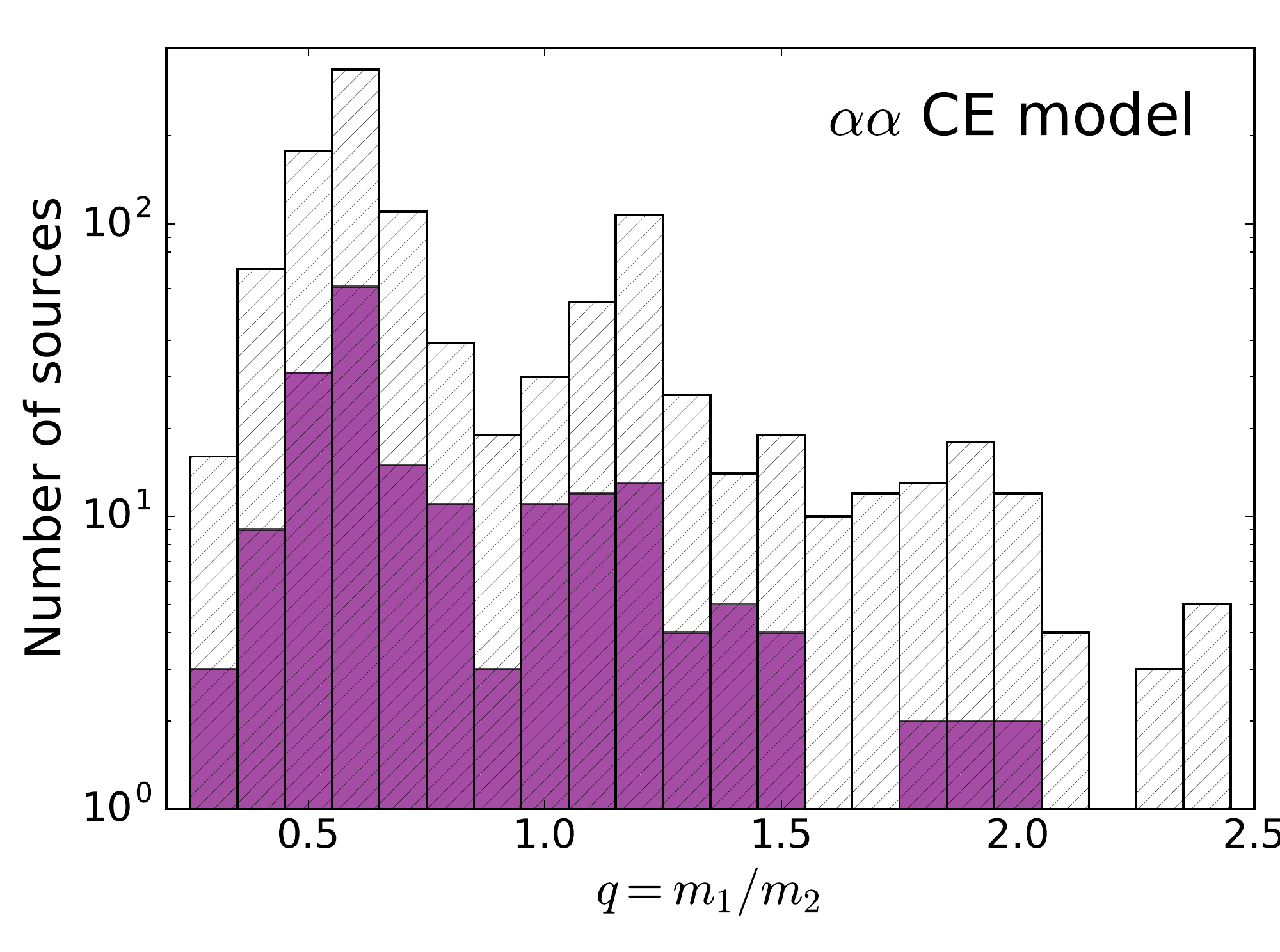}
	 \includegraphics[width=0.4\textwidth]{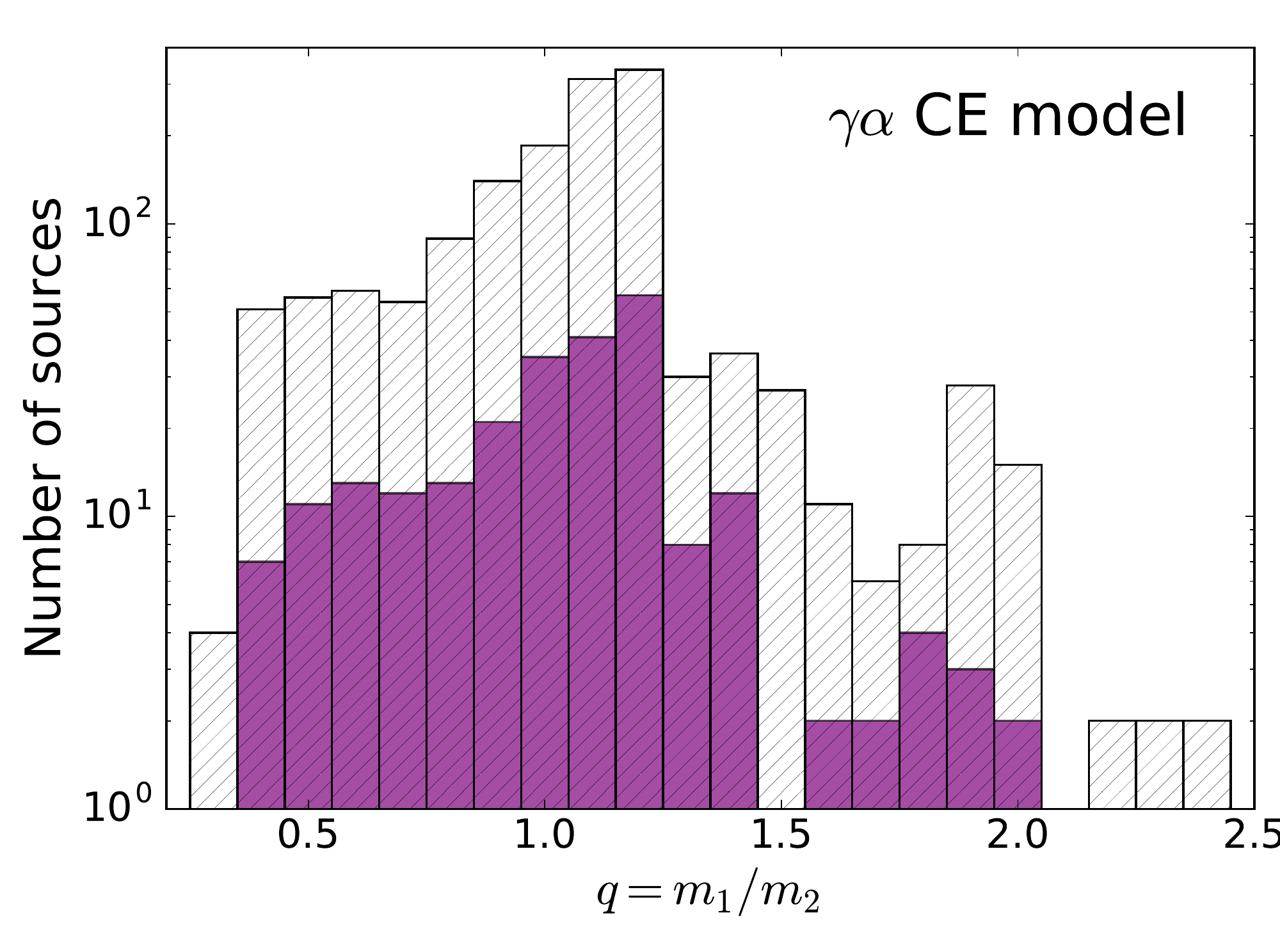}
         \caption{Number of detected sources as a function of binary mass ratio $q=m_1/m_2$ for the two different CE scenarios. The colour coding is the same as Fig.6.}
       \label{fig:7}
\end{figure}
%%%%%%%%%%%%%%%%%%%%%%%%%%%%%%%%%%%

%%%%%%%%%%%%%%%%%%%%%%%%%%%%%%%%%%%
\begin{figure*}
        \centering
	 \includegraphics[width=0.41\textwidth]{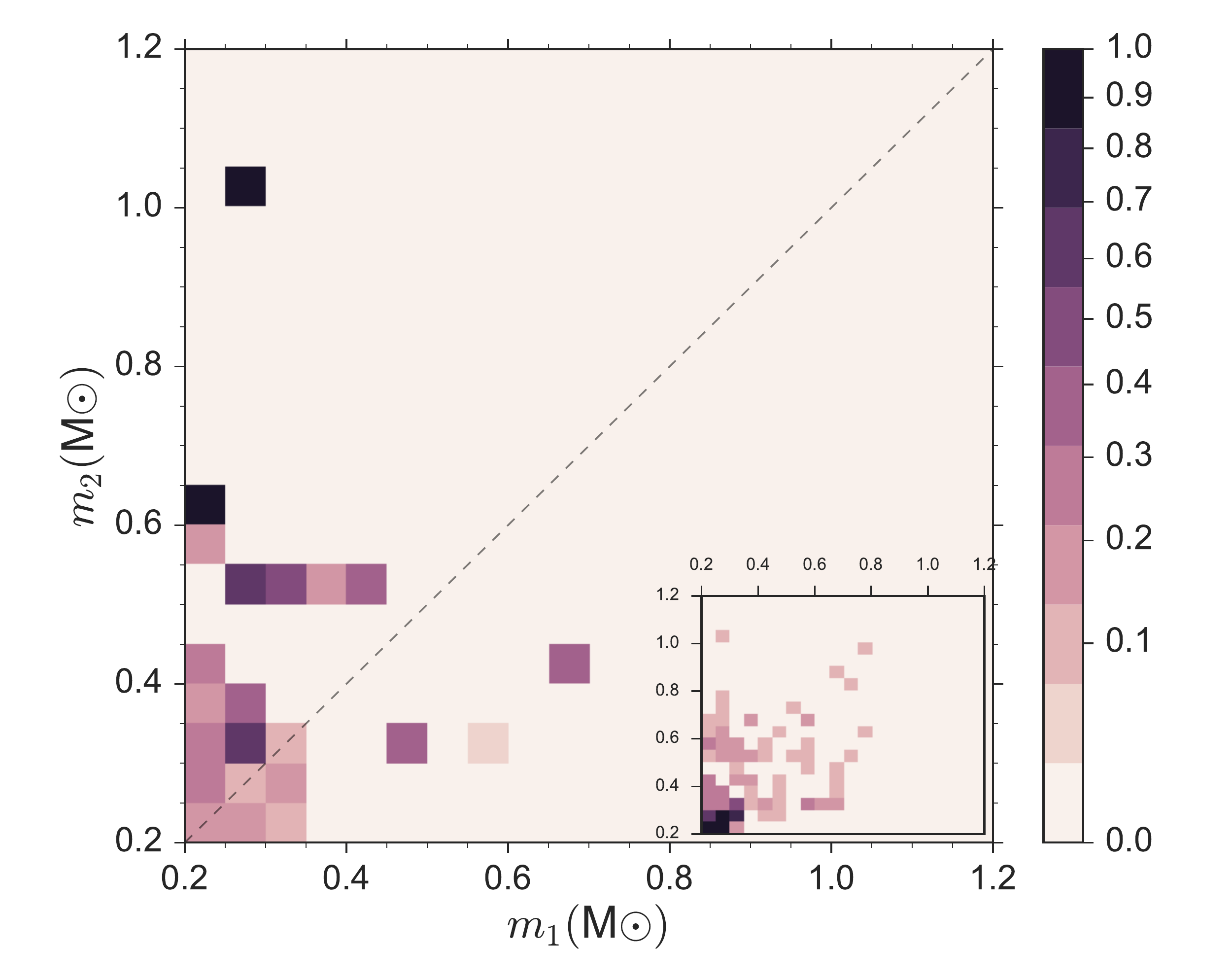}
	 \includegraphics[width=0.41\textwidth]{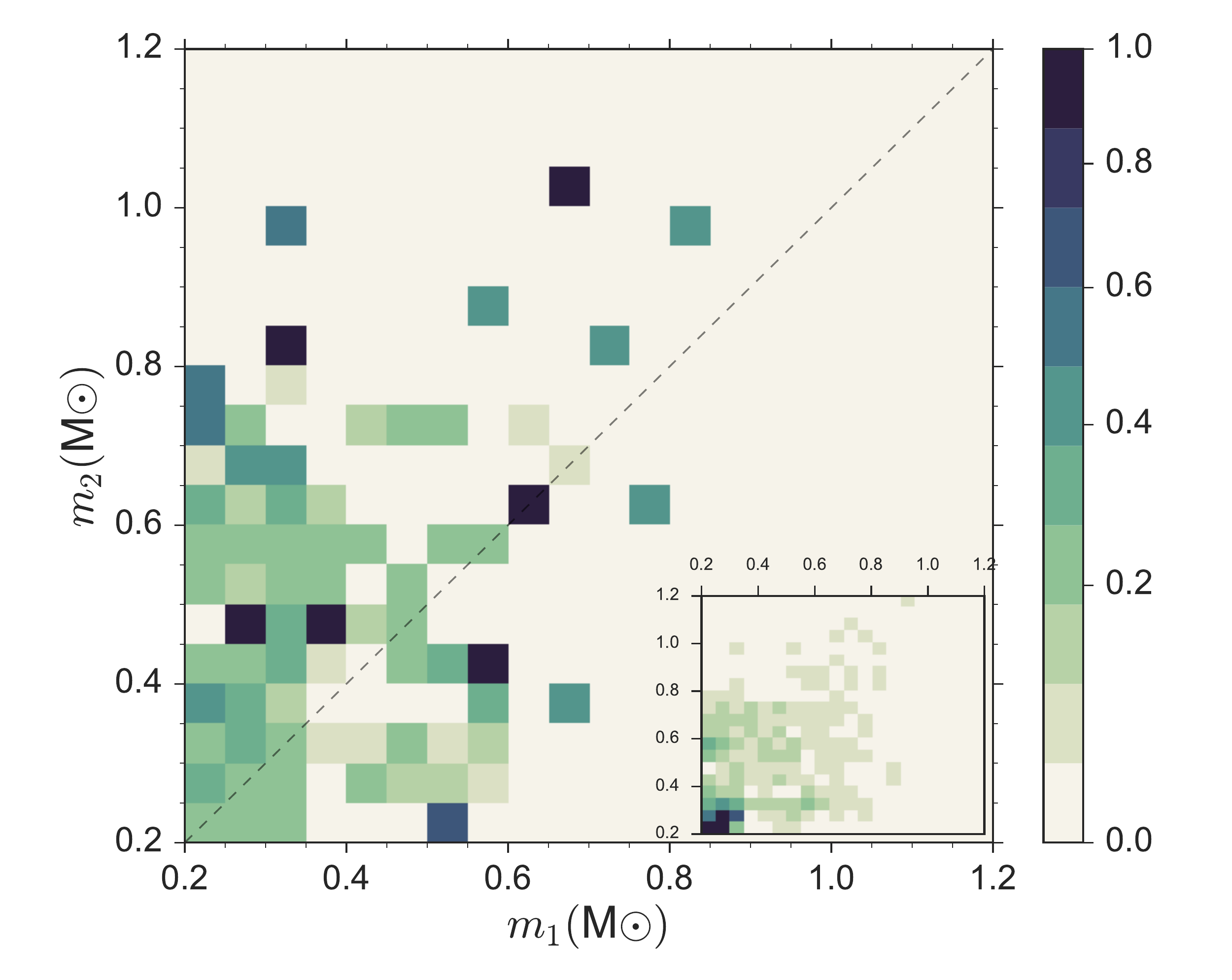}
	 \includegraphics[width=0.41\textwidth]{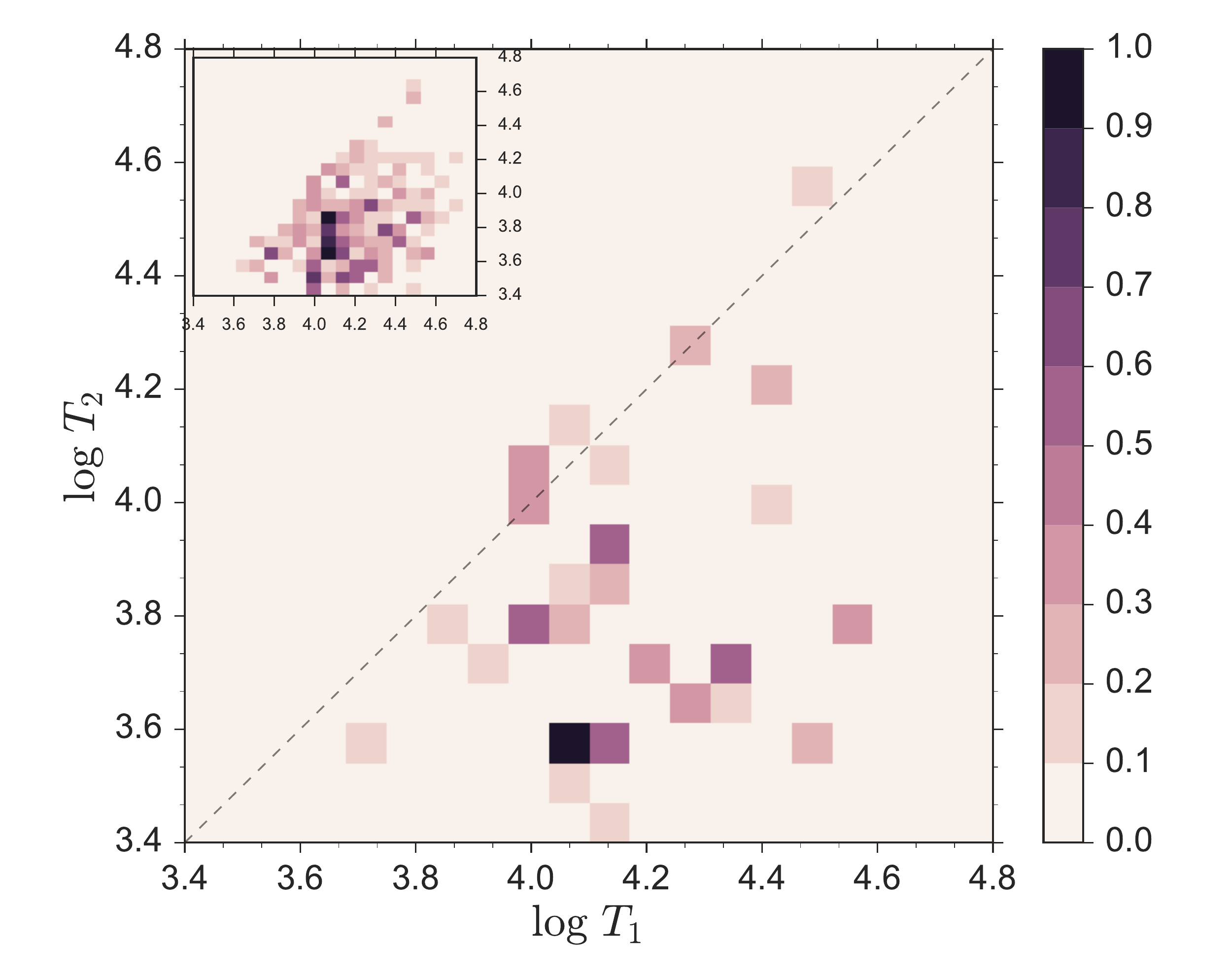}
	 \includegraphics[width=0.41\textwidth]{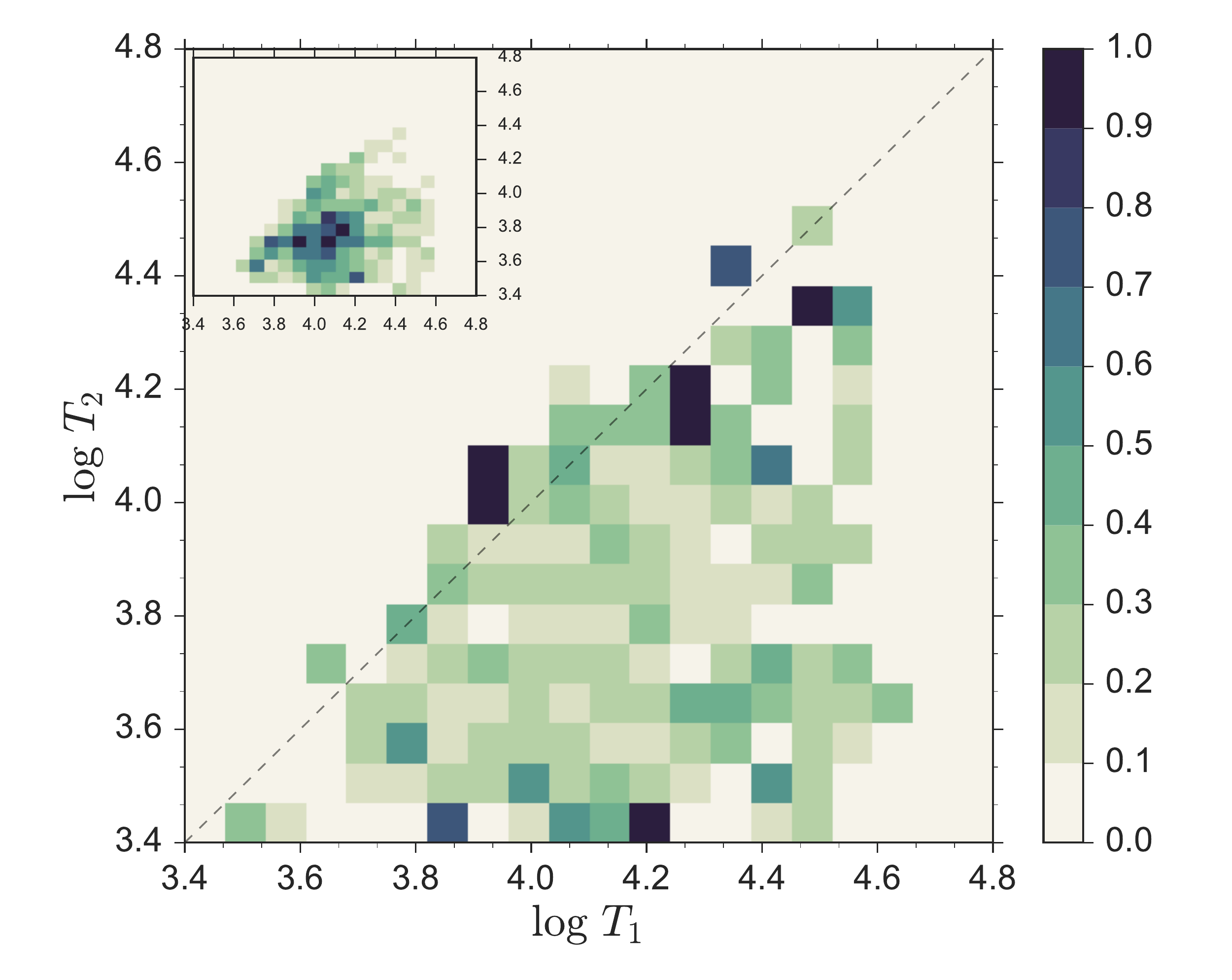}
	 \includegraphics[width=0.41\textwidth]{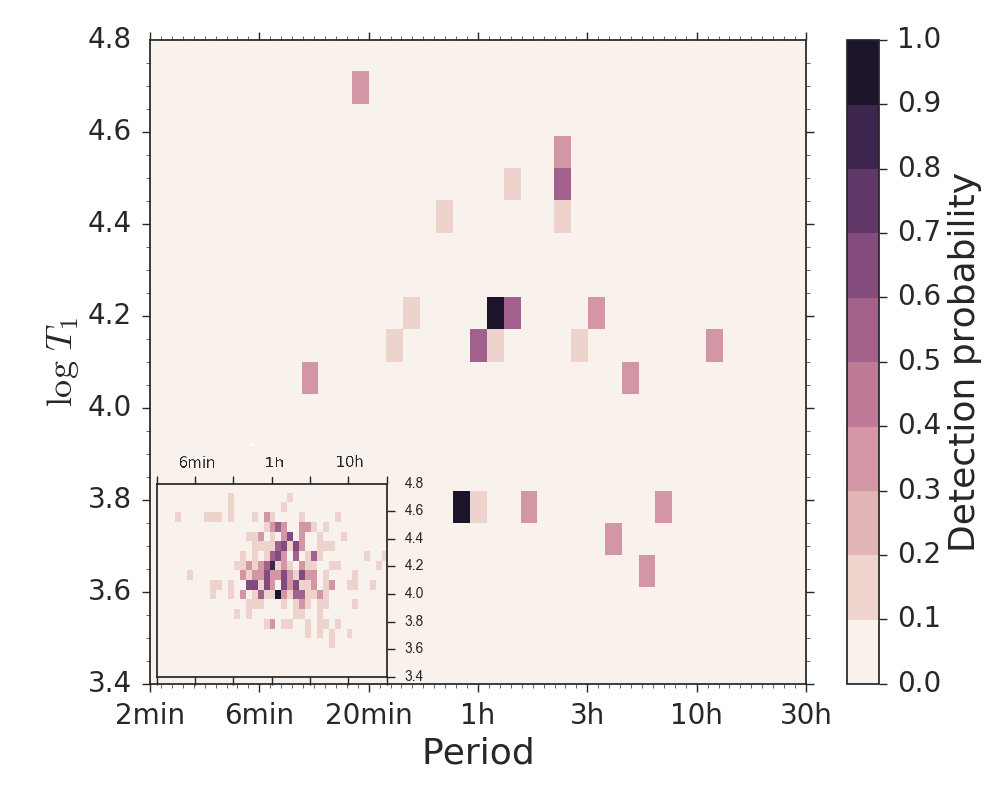}
	 \includegraphics[width=0.41\textwidth]{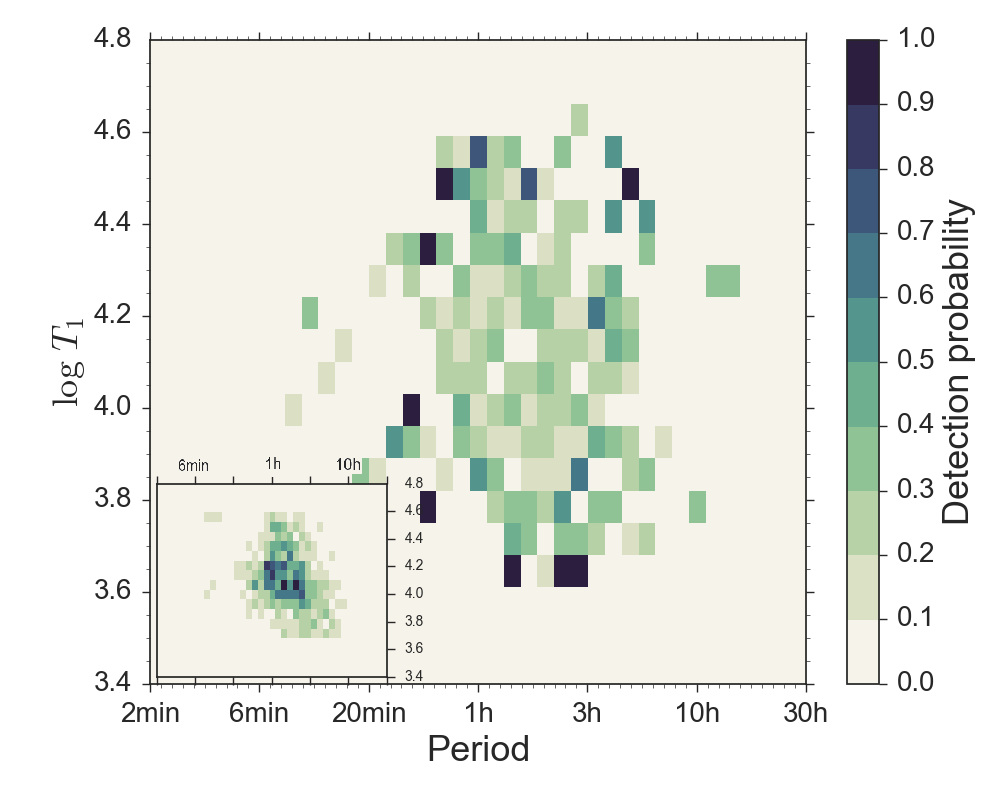}
	 \includegraphics[width=0.41\textwidth]{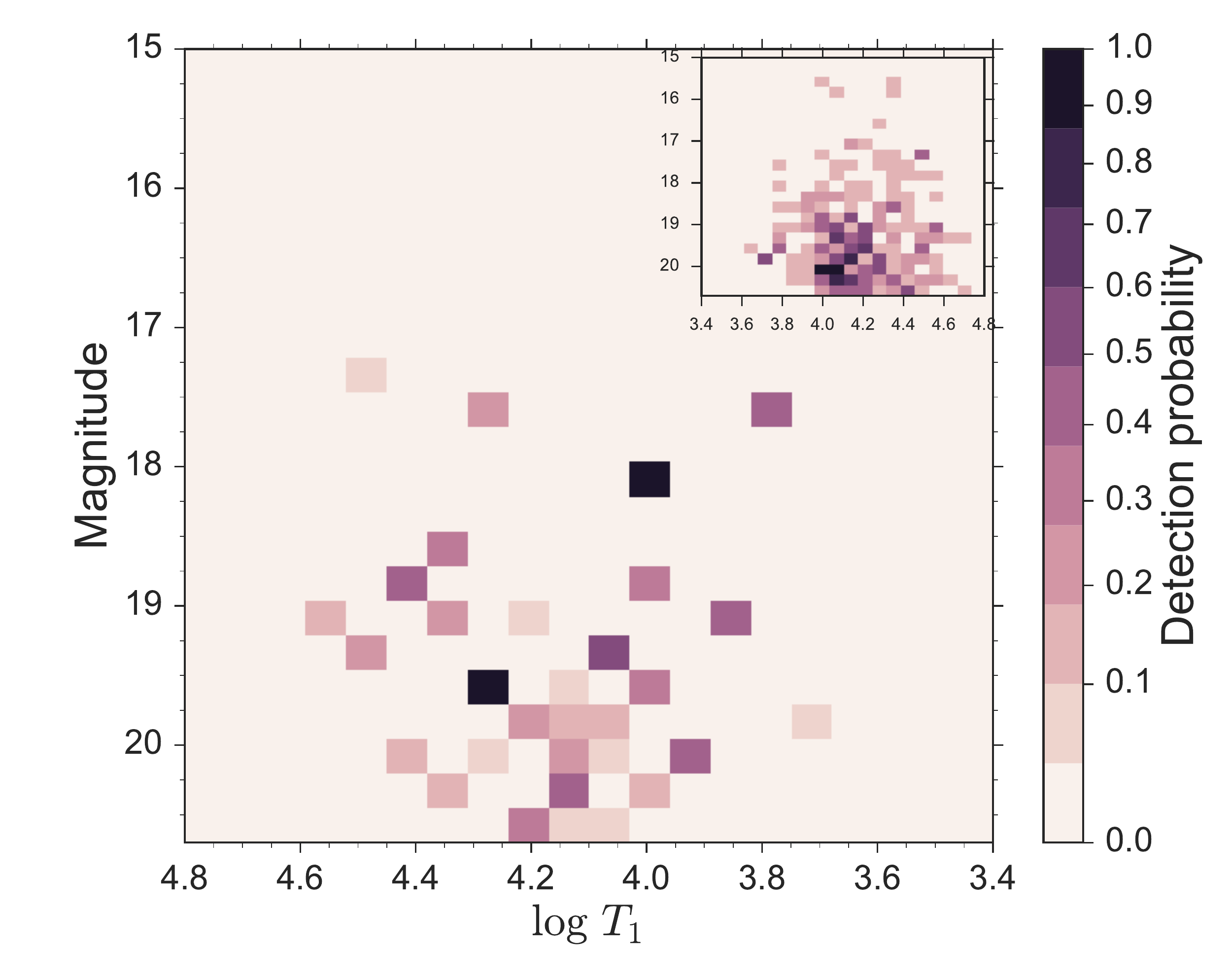}
	 \includegraphics[width=0.41\textwidth]{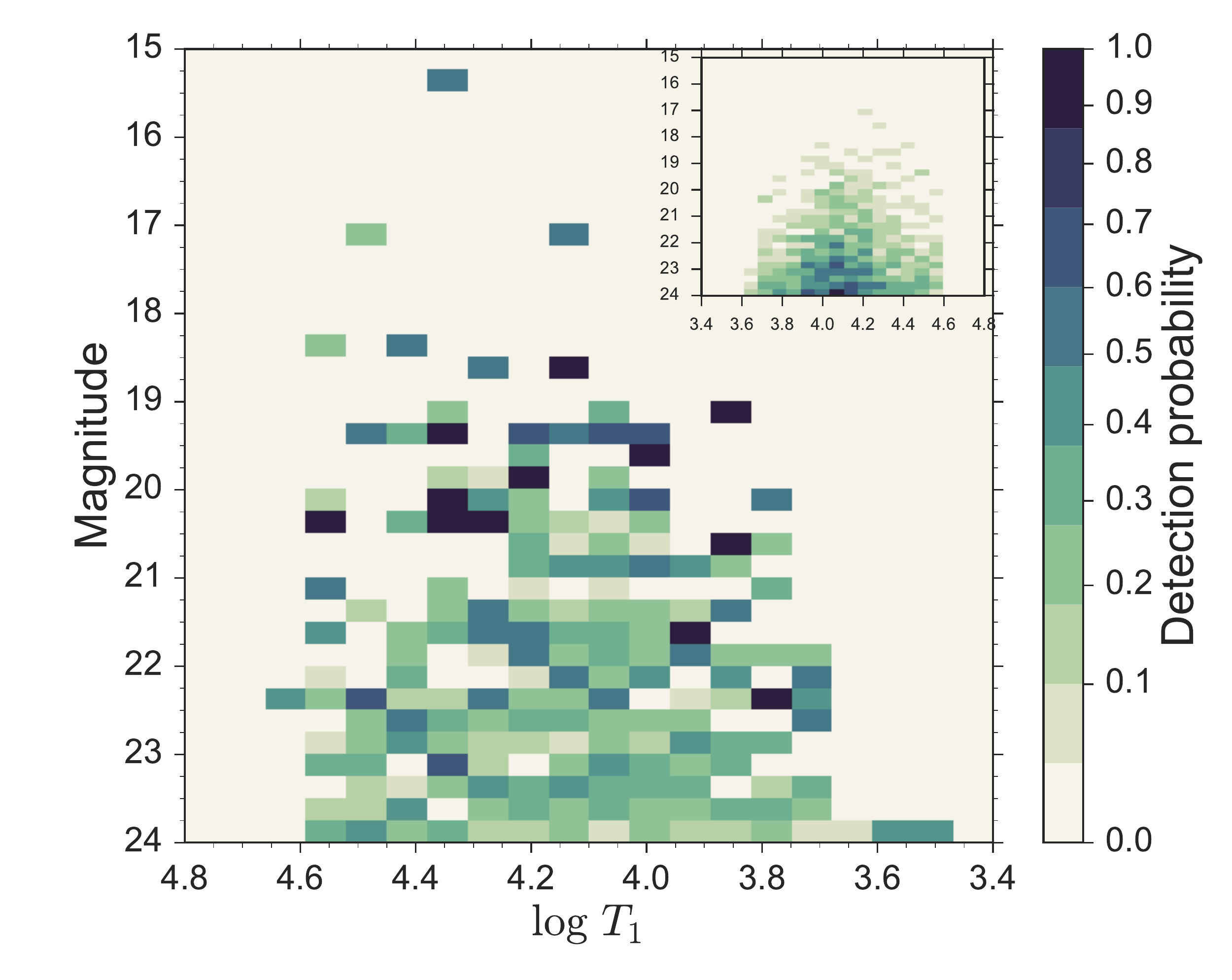}
         \caption{Contour plot for the number of Gaia and LSST detections as a function of binary parameters. We show all the systems formed via the $\gamma \alpha$ scenario weighed by the probability of being detected. The respective inserts represent the distribution of all the systems with a non-zero detection probability. The color indicates the number of detected sources: purple palette for Gaia and green palette for LSST.}
       \label{fig:8}
\end{figure*}
%%%%%%%%%%%%%%%%%%%%%%%%%%%%%%%%%%%%%%

In Fig. 7 we show the fractions of detections formed via the $\gamma \alpha$ scenario for Gaia (purple histogram) and the LSST (hatched histogram) as a function of magnitude, orbital period and binary mass fraction $q=m_1/m_2$.
These fractions are defined as a number of detected sources over the number of binaries of the instrument input population per bin.
For both instruments the fraction of detections drops with increasing orbital periods (top panel).
Note, that the fraction of detected sources by Gaia  > 0.2 even for P > 24 h due to a non-uniform sampling of the light curves because of the peculiar orbital motion of the satellite.
The fraction of detected sources for both instruments decreases with the magnitude from $\sim$1 to 0.5-0.4. 
In particular, we want to draw attention to the fractions of detections as a function of the binary mass ratio (bottom panel of Fig. 7).
Both instruments will be sensitive to binaries with $q>1$, i.e. systems with more massive primaries.
By definition the primary is the brightest WD (and consequently the biggest) of the pair, so a wider range of inclination angles is allowed for these systems in order to be detected as eclipsing sources, and thus they are more likely to be detected.
In our simulation these systems are typically formed via stable mass transfer. 

Figure 8 illustrates the number of detected sources as a function of the mass ratio: top panel for the $\alpha \alpha$ and bottom panel for the $\gamma \alpha$ CE model.
The two distributions are different: the population formed by the $\alpha \alpha$ model shows a prominent peak around $q \sim 0.5$, while the population formed with $\gamma \alpha$ peaks at $q \sim 1$.
Despite the $\gamma \alpha$ CE prescription being designed to match the observed DWD binaries \citep{Nelemans2000, Nelemans2005}, the number of currently known sources is too low to privilege it with respect to the $\alpha \alpha$ CE model.
However, Fig. 8 shows that the Gaia sample has the potential to shed light on the  nature of the CE phase and physical process that triggers it in DWD progenitor systems, since one can already see the difference between the two models by comparing the upper and lower purple histograms.  

In Fig. 9 we illustrate some of the properties of Gaia and LSST detections formed by the $\gamma \alpha$ scenario in different 2D parameter spaces where each source is weighed by probability of being detected; the inserts represent the respective distributions of the sources with a non-zero probability of being detected with an equal weigh.
The detected population will consist of binaries with secondaries typically more massive than primaries.
The majority of known DWDs were discovered by ELM survey, designed to search for extremely low mass primaries, thus new eclipsing binaries detected by Gaia and LSST will extend  this parameter space to binaries with more massive primaries.
Moreover,  the detected population will have primaries hotter than secondaries, therefore it will be difficult to determine directly the properties of the secondaries.
For completeness in the bottom panels of Fig. 9 we represent the distribution of the detected sources in period-temperature and temperature-magnitude space, useful for planning of the spectroscopic follow-up of these sources.

%%%%%%%%%%%%%%%%%%%%%%%%%%%%%%%%%%%%%%%%%%%%%%%%%%%%%%%%%%%%%
%%%%%%%%%%%%%%%%%%%%%%%%%%%%%%%%%%%%%%%%%%%%%%%%%%%%%%%%%%%%%

\section{GW detection}

In this section we focus our attention on DWD binaries as GW sources. 
First, we recall some basic formulae for the estimation of the GW signal and
we refer to Appendix A the summary of the method, developed by \citet{CL2003}, used in this work to simulate the LISA instrument response.
Then, we estimate the signal-to-noise ratio (SNR) of currently observed DWD binaries to verify our procedure.
The following step is to calculate the SNR for all our synthetic binaries to identify those with the highest SNR.
Finally, we compare our result with previous works \citep{Nelemans2004,Nissanke2012}, that are based on a different Galactic model population \cite[for detailed analysis of the deferences between their model and ours see][Sect. 2]{Toonen2012}.

LISA is a space-based gravitational wave interferometer,
conceived as a set of three spacecrafts in an equilateral triangle constellation of a few million km per side.
Such spacecraft separation sets the sensitivity range of the instrument from about 0.1 to 100 mHz and will allow the detection of Galactic and extra-Galactic sources, among which thousands will be DWD binaries \citep{LISA2017}.
The detector's center-of-mass will follow a circular heliocentric trajectory, trailing 22$^{\circ}$ behind the Earth and maintaining a  60$^{\circ}$ inclination between the plane of the detector and the ecliptic plane.
As the reference LISA configuration used for this work, we adopt the LISA Mission Concept recently submitted as a response to the ESA call for L3 missions (hereafter ESACall v1.1).
The ESACall v1.1 is a three-arm configuration\footnote{Note that each arm corresponds to two laser links between spacecrafts, so that a three-arm detector consists of six links in total.} with $2.5 \times 10^6$ km arm length instead of $5 \times 10^6$ km arm length as in the original LISA project \citep[see, e.g.,][]{2007AAS...21114001P}.
The sensitivity of the ESACall v1.1 configuration, is based on the latest results from the LISA Pathfinder mission \citep{LPF2016}, a precursor mission designed to test the technologies needed for the laser interferometry in space.
It is represented in Fig. 10 \citep{LISA2017}.

As pointed out by several authors, at frequencies below a few mHz the expected number of Galactic binaries (in particular, detached DWDs) per frequency bin ($\triangle f = 1/T_{\rm obs}$, where $T_{\rm obs}$ is the total observation time) is so large that these binaries will form an unresolvable foreground signal in the detector \citep[e.g.,][]{2007AAS...21114001P, Amaro-Seoane2012}.
Figure 10 illustrates the foreground level from detached DWDs computed by using our model population \citep{LISA2017}, and its evolution with time from 0.5 to 10 yr of observation.

%%%%%%%%%%%%%%%%%%%%%%%%%%%%%%%%%%%
\begin{figure}
        \centering
	 \includegraphics[width=0.45\textwidth]{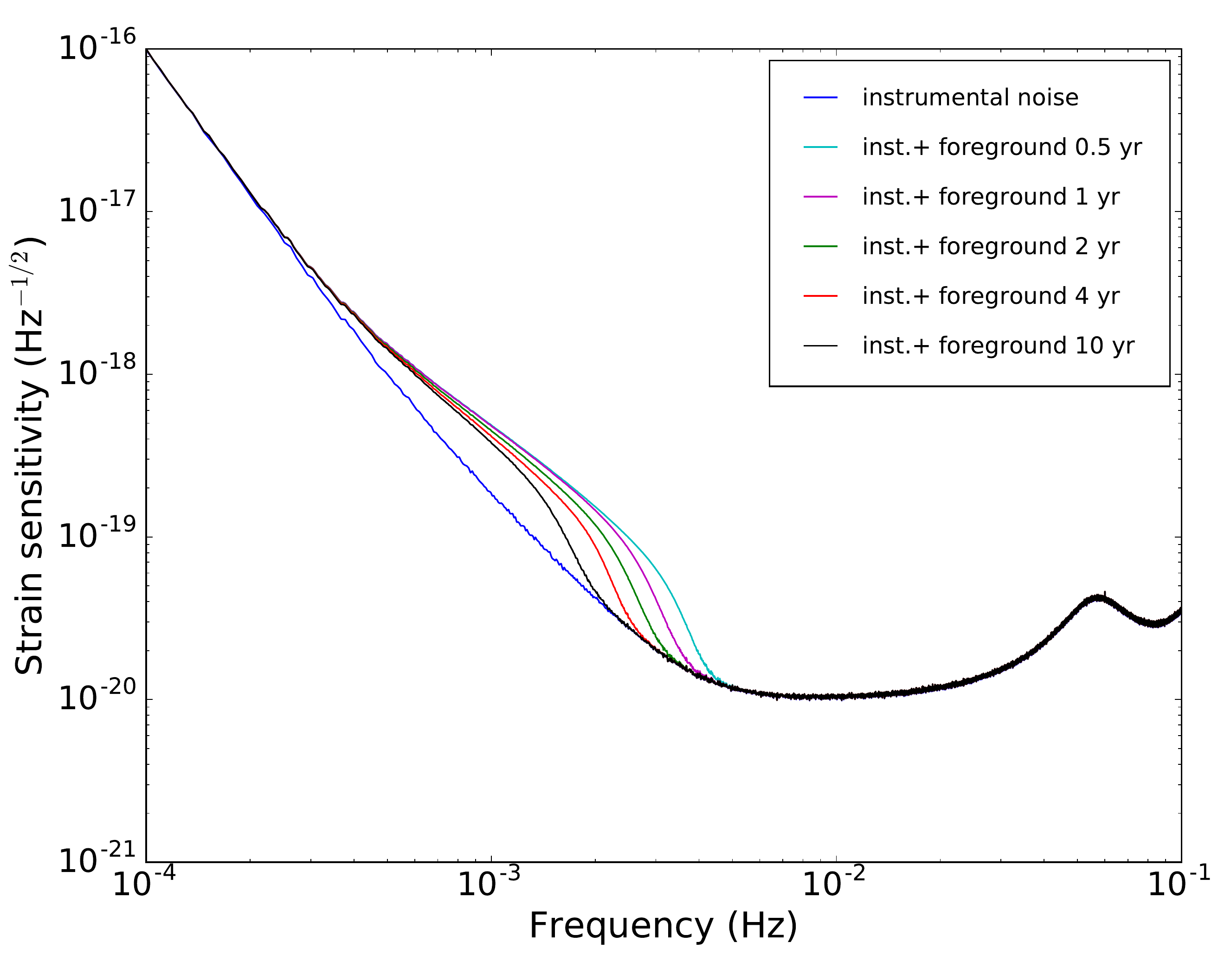}
         \caption{LISA ESACall v1.1 sky-averaged sensitivity \citep{LISA2017} due to the instrumental noise only and due to the instrumental noise plus Galactic foreground from DWD binaries after 6 mouths, 1, 2, 4 and 10 years of observations.}
       \label{fig:9}
\end{figure}

\begin{figure*}
        \centering
	 \includegraphics[width=0.9\textwidth]{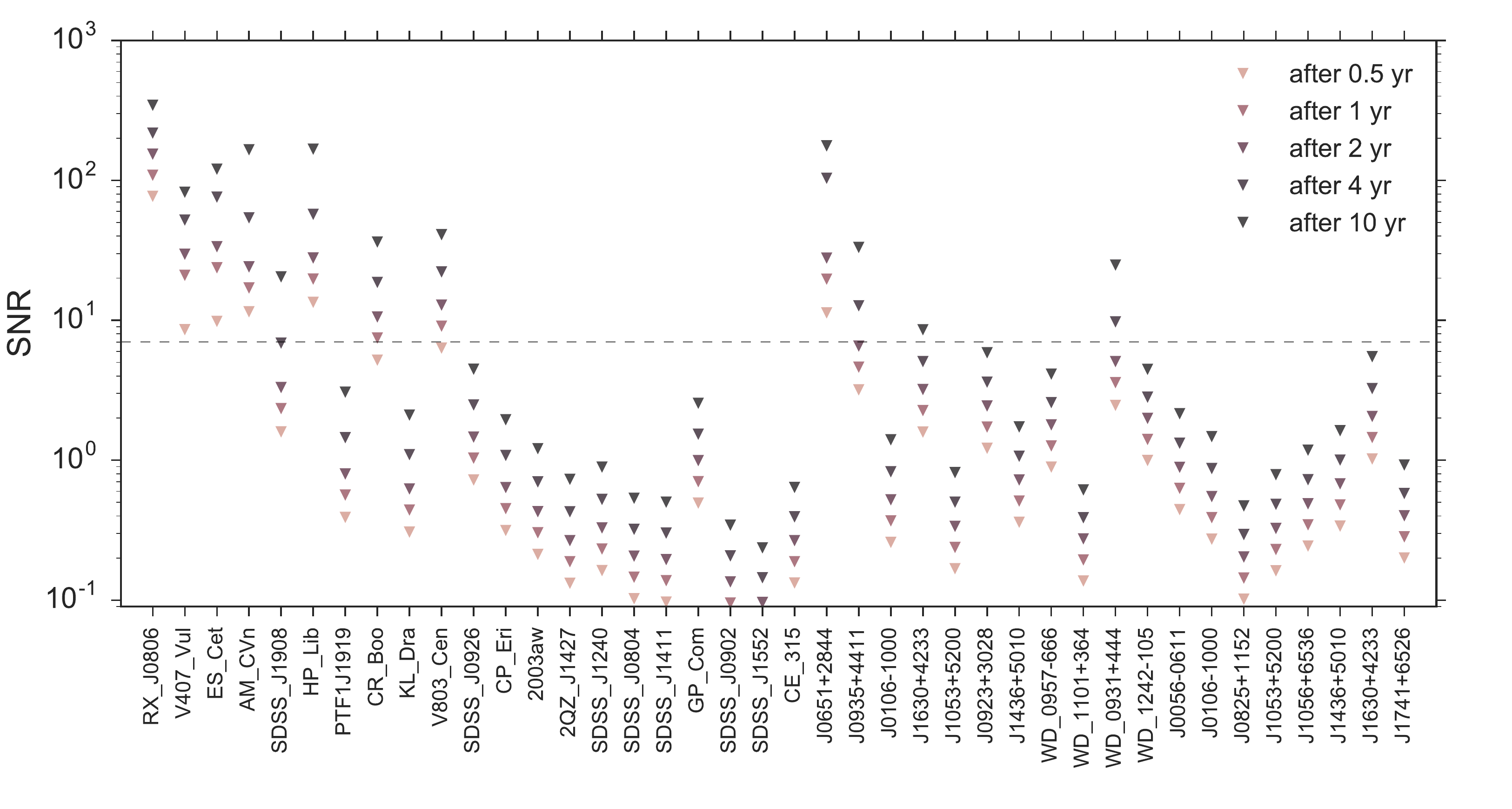}
         \caption{SNR evolution with time for a sample of LISA verification binaries. The black dashed line corresponds to SNR = 7.}
       \label{fig:10}
\end{figure*}
%%%%%%%%%%%%%%%%%%%%%%%%%%%%%%%%%%%

%%%%%%%%%%%%%%%%%%%%%%%%%%%%%%%%%%%%%%%%%%%%%%%%%%%%%%%%%%%%%%

\subsection{GW signal from DWDs}

The great majority of Galactic DWD binaries can be well described using Newtonian dynamics of circular orbits.
The gravitational waves they produce can be computed using the quadrupole approximation \citep[see, e.g.,][]{Landau, Peters1963}. 
Considering that the timescale on which DWDs typically evolve  (> Myr) is much greater than the  lifetime of the LISA mission ($\sim$ yr), they can be treated as monochromatic sources emitting at the frequency $f_{\rm s} = 2/P$.
In this approximation the GW signal emitted by a binary is given by a combination of the two polarizations:
\begin{equation}
	h_+ (t)=  \frac{2(G {\cal M})^{5/3}(\pi f_{\rm s})^{2/3}}{ c^4 d}(1 + \cos^2 i)\cos 2\Phi(t), 
\end{equation}
\begin{equation}
	h_{\times} (t)=  - \frac{4(G {\cal M})^{5/3}(\pi f_{\rm s})^{2/3}}{ c^4 d} \cos i \sin 2\Phi(t),
\end{equation}
where ${\cal M} = (m_1m_2)^{3/5}(m_1+m_2)^{-1/5}$ is the chirp mass of the system, $\Phi(t) = \Phi_0 + \pi f_{\rm s} t$ is the orbital phase, and $i$ is the inclination of the binary orbital plane with respect to the line of sight.

In the low frequency limit ($f \ll c/2\pi L \sim 20$ mHz, where $L=2.5$ Mkm is the detector's arm length) the GW signal as measured  by the detector can be expressed as
\begin{equation}
h(t) = F_+ h_+(t) +F_{\times} h_{\times}(t),
\end{equation}
where $F_+$ and $F_{\times}$ are the detector pattern functions (see Appendix A), that depends on the source location in the sky, its orientation with respect to the detector and the detector configuration.
For a monochromatic periodic source the signal-to-noise ratio (SNR) can be written as \citep[][Eq.(7.129)]{MaggioreGW}:
\begin{equation}
\left( \frac{S}{N} \right)^2 = 4 \int_{0}^{\infty} df  \frac{|\tilde{h}(f)^2|}{S_{\rm n}(f)} = \frac{{\cal A}^{2} T_{\rm obs}}{S_{\rm n}(f_{\rm s})}, 
\end{equation}
where $\tilde{h}(f)$ is the Fourier transform of $h(t)$, $S_n(f_{\rm s})$ is the noise spectral density of the instrument (Fig. 9) at $f_{\rm s}$, the signal amplitude is
\begin{equation}
{\cal A} = {\left[ h_+^2 F_+^2 (t) + h_{\times}^2 F_{\times}^2 (t)\right]}^{1/2},
\end{equation}
and $T_{\rm obs}$ is equal to the mission life time for DWD binaries.
Note that eq.(10) represents a general result for a generic two arm interferometer.
To compute a realistic LISA response we followed the frequency domain method developed by \citet{CL2003} summarized in Appendix A.
This method represent a fast algorithm for the computation of the orbit average amplitude of the signal $\left<{\cal A}\right>$ (see the resulting equation eq.(A12) of Appendix A).

%%%%%%%%%%%%%%%%%%%%%%%%%%%%%%%%%%%%%%%%%%%%%%%%%%%%%%%%%%%

\subsection{Results}

%%%%%%%%%%%%%%%%%%%%%%%%%%%%%%
\begin{table}
\begin{centering}
\caption{Total number of individually resolved DWDs with SNR > 7 for the LISA ESACallv1.1 mission configuration.}
\centering
 \begin{tabular}{c | c | c | c | c | c}
 \hline
    CE model & 6 m & 1 yr & 2 yr & 4 yr & 10 yr \\ \hline \hline
   $\alpha \alpha$ & 6185 & 9808 & 16360 & 24482 & 44349\\ \hline
   $\gamma \alpha$ & 7125 & 11385& 18330& 25754& 52045\\ \hline
  \end{tabular}
\end{centering}
\end{table}
%%%%%%%%%%%%%%%%%%%%%%%%%%%%%

\begin{figure*}
        \centering
	 \includegraphics[width=0.45\textwidth]{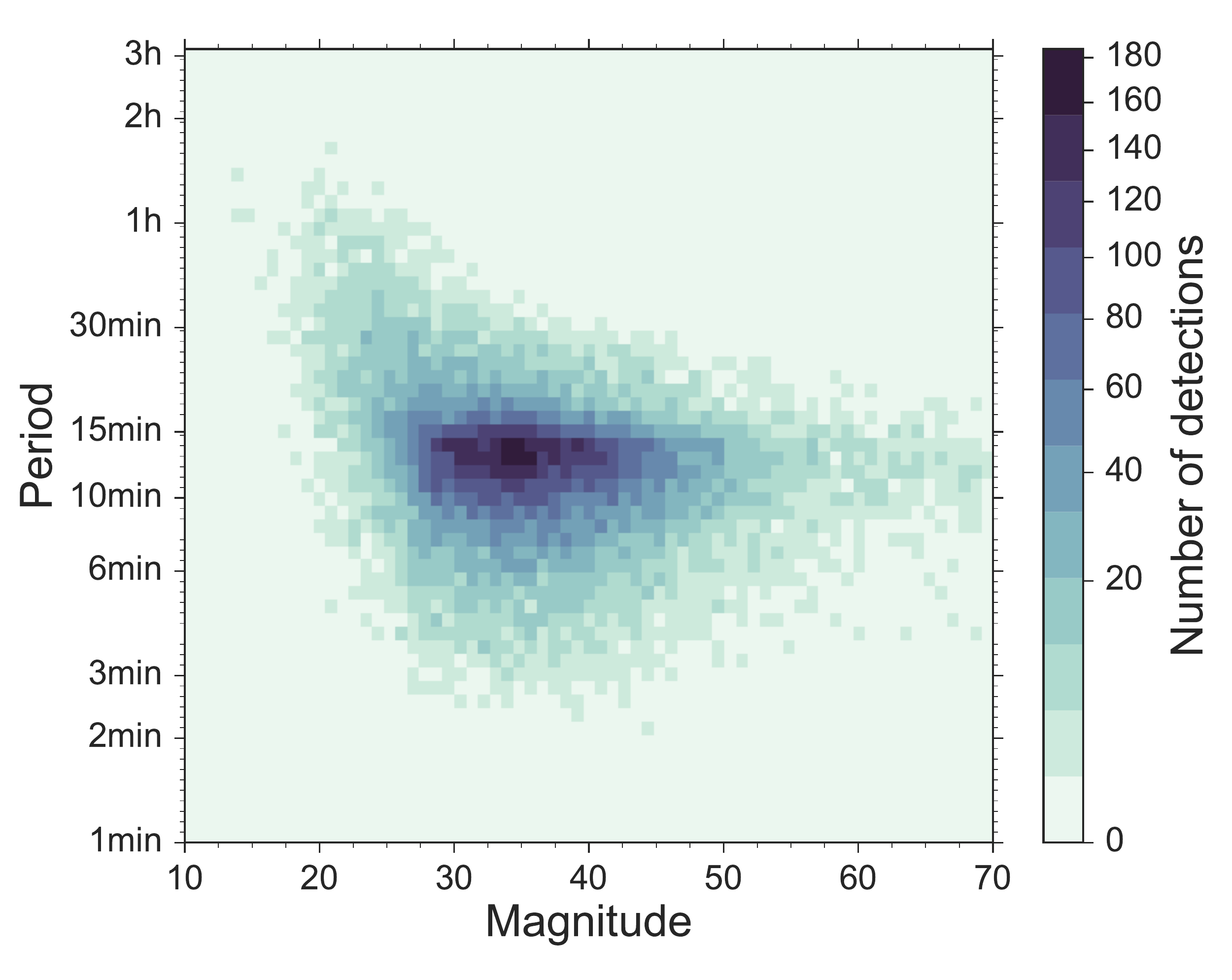}
	 \includegraphics[width=0.45\textwidth]{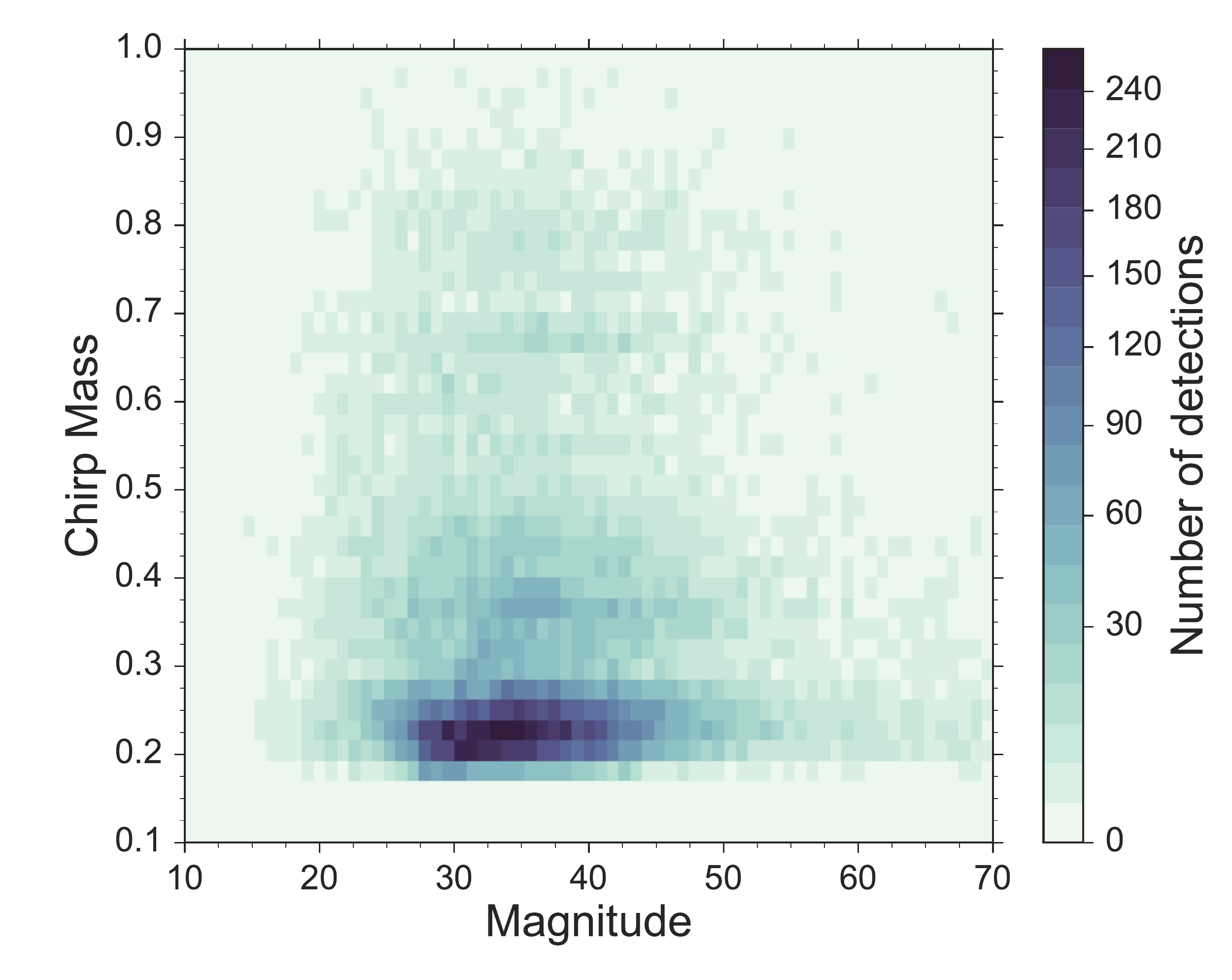}
	 \includegraphics[width=0.45\textwidth]{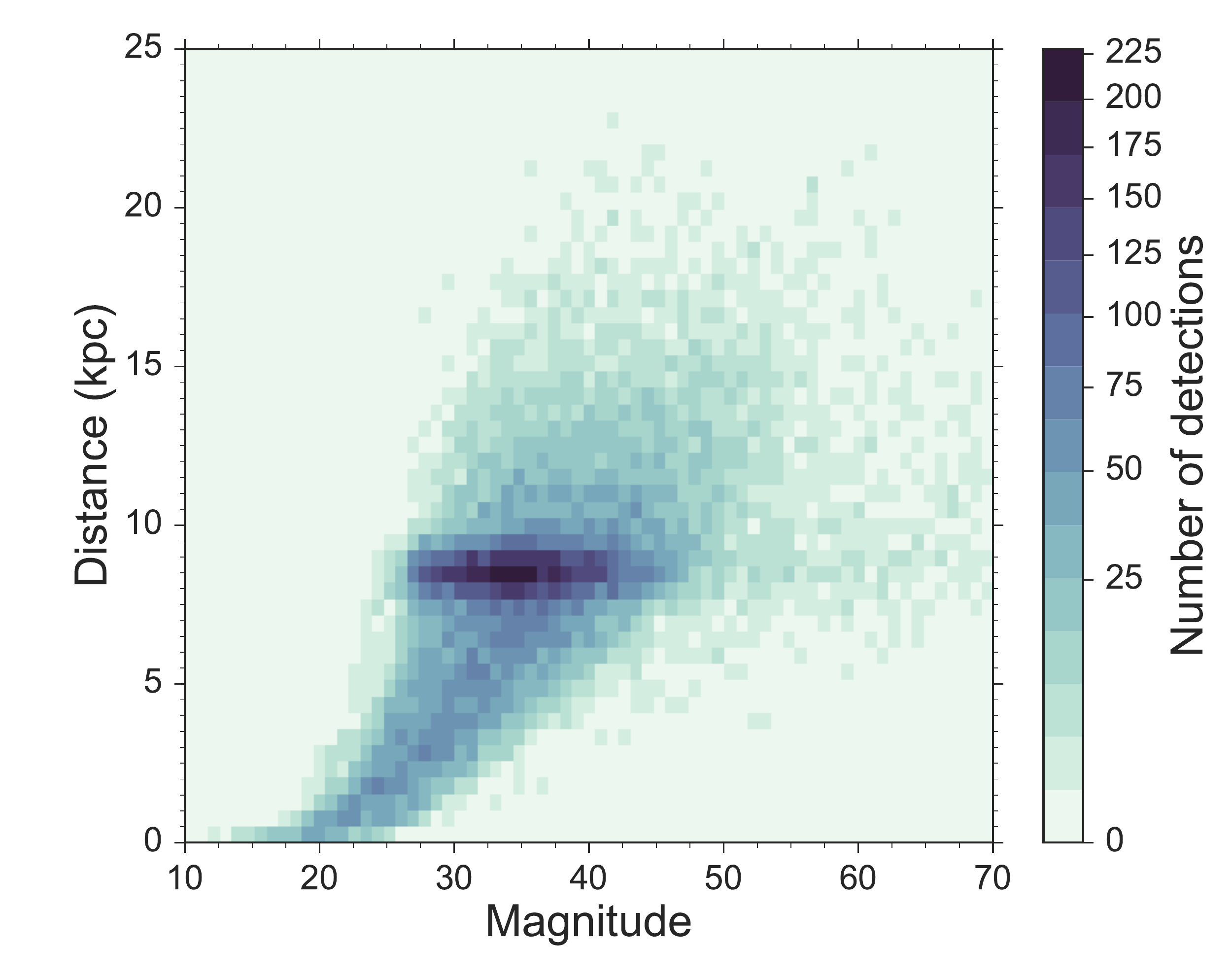}
	 \includegraphics[width=0.45\textwidth]{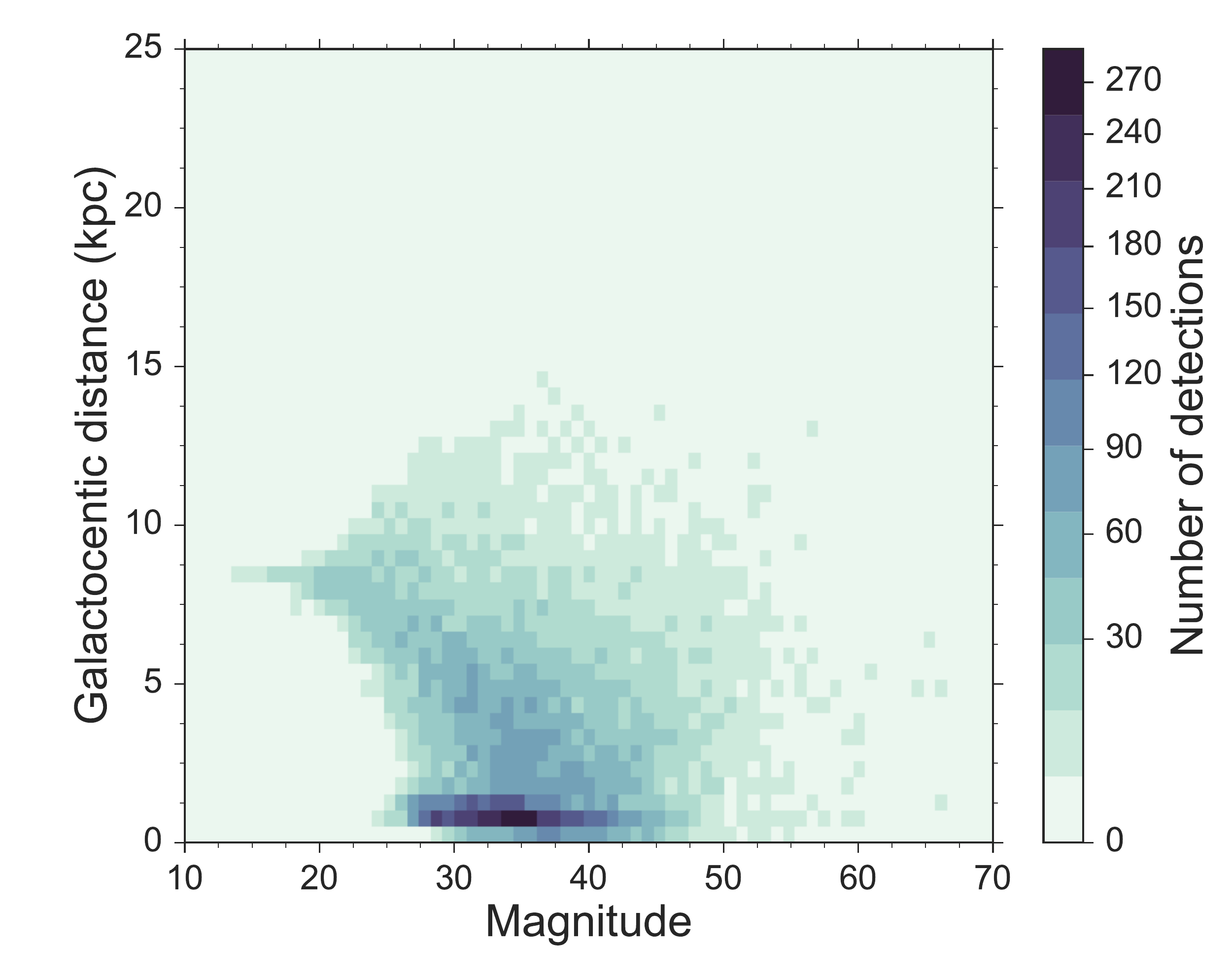}
         \caption{2D histograms of the number of LISA detections with SNR > 7. The colour indicates the number of detected sources.}
       \label{fig:10}
\end{figure*}

To test our method we consider the sample of the so-called verification binaries. 
These are well known ultra-compact binaries (mostly detached DWDs and AM CVns), that are expected to be bright in the LISA band. 
Consequently, they represent guaranteed sources for the mission. 
Some of these binaries will be detected in a short period after the beginning of the mission ($\sim$ few months), and thus can be used to verify the performance of the instrument \citep[e.g.][]{Stroeer2006}.
The verification binary parameters are listed in Table B1.

For each verification binary we compute the SNR by plugging into eq.(10) the orbit-averaged amplitude as seen by LISA (eq.A12).
Our results for 1 yr mission lifetime are reported in Table B1.
We find that 8 of the 58 verification binaries reported in Table B1 have SNR > 7 within the first year of observation, and 10 within the nominal mission life time of 4 years.
These results are in agreement with the full time domain LISA simulation \citep[A. Petiteau private communication, see also][]{LISA2017}. 
Figure 10 also illustrates how the SNR grows progressively with mission duration.

We compute the total number of resolved binaries in our model DWD population.
The polarization angle and the initial orbital phase (respectively $\psi$ and $\phi_0$), that are not provided directly by our population synthesis simulation, are randomize assuming uniform distribution over the interval of their definition $[0, 2\pi)$.
The result for the two formation scenarios and different mission durations are reported in Table 3.
The number of individually resolved DWDs for the LISA ESACallv1.1 configuration with SNR > 7 are $\sim 10-11 \times 10^3$ for 1 year and $24.5 - 25.7 \times 10^3$ for 4 year of mission.
Our result is compatible with those obtained by the Gravitational Observatory Advisory Team (GOAT)\footnote{http://sci.esa.int/jump.cfm?oid=57910}, \citet{Shah2012} and \citet{Nissanke2012}, based on Galactic population from \citet{Nelemans2004}, when considering a different arm length.

In Fig. 12 we show some of the properties of LISA detections expected in the $\gamma \alpha$ CE model.
From comparison between Fig. 12 and Fig. 6 it is evident that LISA will see binaries that are non accessible to EM detectors (virtually down to magnitude 70).
LISA detections will have periods ranging between 2 min and 2h, and chirp masses up to 1 M$_{\odot}$.
Remarkably, unaffected by extinction, LISA will see binaries throughout the Galaxy up to distances comparable with the extension of the Galactic disc (Fig. 12 bottom panels).
In particular, LISA will detect DWDs even beyond the Galactic center, that is impossible with EM observations.
Figure 12 shows that the most of the detections comes from Galactic bulge (i.e. at Galactocentric distance close to 0).

%%%%%%%%%%%%%%%%%%%%%%%%%%%%%%%%%%%%%%%%%%%%%%%%%%%%%%%%%%%%%%%
%%%%%%%%%%%%%%%%%%%%%%%%%%%%%%%%%%%%%%%%%%%%%%%%%%%%%%%%%%%%%%%

\section{The combined EM \& GW sample}

In the two previous sections we showed that the expected number of DWD detections through EM and GW radiation within next two decades is significant.
So far GW studies  focused on currently known Galactic binaries or on the future EM follow up of these sources, ignoring the fact that revolutionary optical surveys such as Gaia and the LSST will be available between now and the LISA launch. 
In this section we want to estimate how many DWDs detected by Gaia and LSST will be bright enough in GWs to be also detected by the LISA.

Starting from the Gaia and LSST samples (Sect. 3) we compute the SNR for the LISA ESACallv1.1 configuration and 4 yr mission lifetime, and we select those with SNR > 7 (see Tab. 4).
We find 13 and 25 combined Gaia and LISA detections respectively for the $\alpha \alpha$ and $\gamma \alpha$ CE models.
Combined LSST and LISA samples are 3-4 times bigger: 50 in the $\alpha \alpha$ formation scenario and 73 for in $\gamma \alpha$ scenario.
This result shows that before the LISA launch we will have at least twice as many guaranteed LISA detections with SNR > 7.
The period of the combined detections will range from a few minutes to 1 hour, and will be on average (as for currently known LISA verification binaries) around 15 min (see Fig 13).
As for the sample of known verification binaries (Table B1), the mass of the primary, secondary, and, consequently the chirp mass of these binaries is not expected to exceed 1 M$\odot$.
Verification binaries provided by Gaia are not expected to be found at distances larger than the already known ones,
while the LSST will double the maximum distance because of its deeper photometric limit (Fig. 13).

Several authors has already pointed out that for these combined detections much more information can be gained compare to either EM or GW can provide alone \citep[see, e.g.,][]{Marsh2011, Shah2012,Shah2014}.
Light curves alone allow the measurement of the orbital period, the inclination angle and the scaled radii of the binary components ($R_1/a$ and $R_2/a$), that, in turn, can be used to determine the binary mass ratio from the mass-radius relationship.
This information combined to the chirp mass determined from the GW data, in principle, permit to estimate single masses of the binary components. 
For monochromatic sources, like DWD binaries, GW data will provide the measurement of the chirp mass in combination with distance (see eqs.(7)-(11)).
The EM measurement of the distances will be crucial to break this degeneracy \citep{Shah2012}.
Furthermore, measuring $\dot{f}$ from GWs (not likely for DWDs and was not taken into account in our simulation) or $\dot{P}$ from eclipse timing has the same effect \citep{Shah2014}.

Furthermore, the EM data can be also used to constrain GW observables and to improve their accuracy.
In fact, there are several correlations between the GW and EM observable quantities: binary inclination, ecliptic latitude and longitude and GW amplitude.
For example, an a priori knowledge of the source sky position and inclination can give an improvement on the measurement of GW amplitude up to a factor of 60 \citep{Shah2013}.
Vice versa, \cite{Sweta2012} showed that small inclination errors from GW data imply that system is eclipsing, consequently this fact can be used for the EM detection of new eclipsing sources.

%%%%%%%%%%%%%%%%%%%%%%%%%%%%%%
\begin{table}
\captionsetup[subfigure]{position=top}
%\small
\centering
\caption{Summary table for the number of detections with Gaia, LSST and LISA. We reported results for the nominal mission life time: 5 yr for Gaia, 10 yr for the LSST and 4 yr for LISA.}
  \subfloat[$\alpha \alpha $ CE model]{%
    \hspace{.5cm}%
\begin{tabular}{|c|c|c|c|}
\hline
     & Gaia & LSST & LISA  \\ \hline
Gaia & 189  & 93   & 13    \\ \hline
LSST & 93   & 1100 & 50    \\ \hline
LISA & 13   & 50   & 24508 \\ \hline
\end{tabular}
\hspace{.5cm}%
}\hspace{1cm}
\subfloat[$\gamma \alpha $ CE model]{%
\hspace{.5cm}%

\begin{tabular}{|c|c|c|c|}
\hline
     & Gaia & LSST & LISA  \\ \hline
Gaia & 246  & 155  & 24    \\ \hline
LSST & 155  & 1457 & 73    \\ \hline
LISA & 24   & 73   & 25735 \\ \hline
\end{tabular}
\hspace{.5cm}
}
\end{table}
%%%%%%%%%%%%%%%%%%%%%%%%%%%%%

%%%%%%%%%%%%%%%%%%%%%%%%%%%%%
\begin{figure}
        \centering
	 \includegraphics[width=0.4\textwidth]{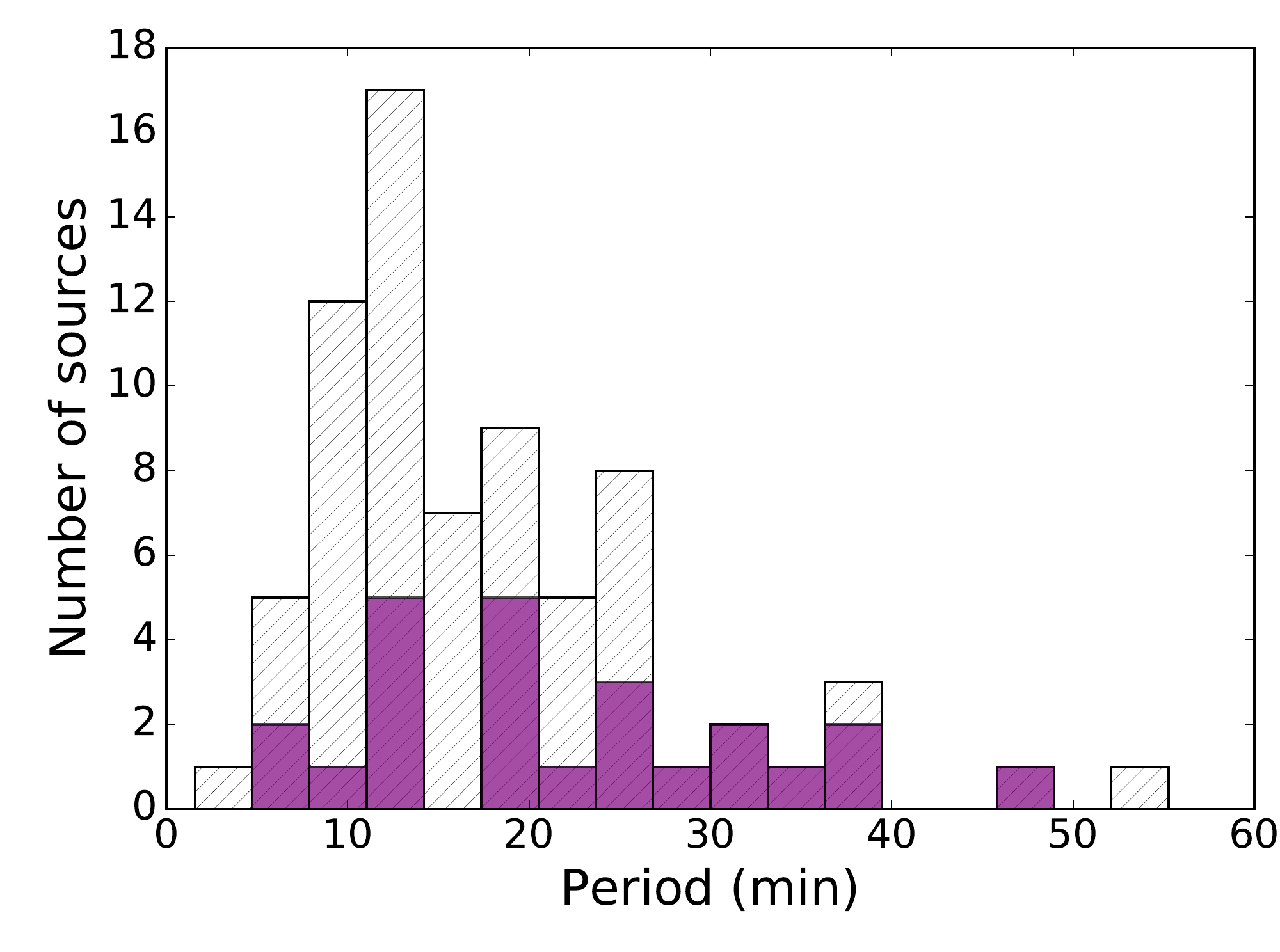}
	 \includegraphics[width=0.4\textwidth]{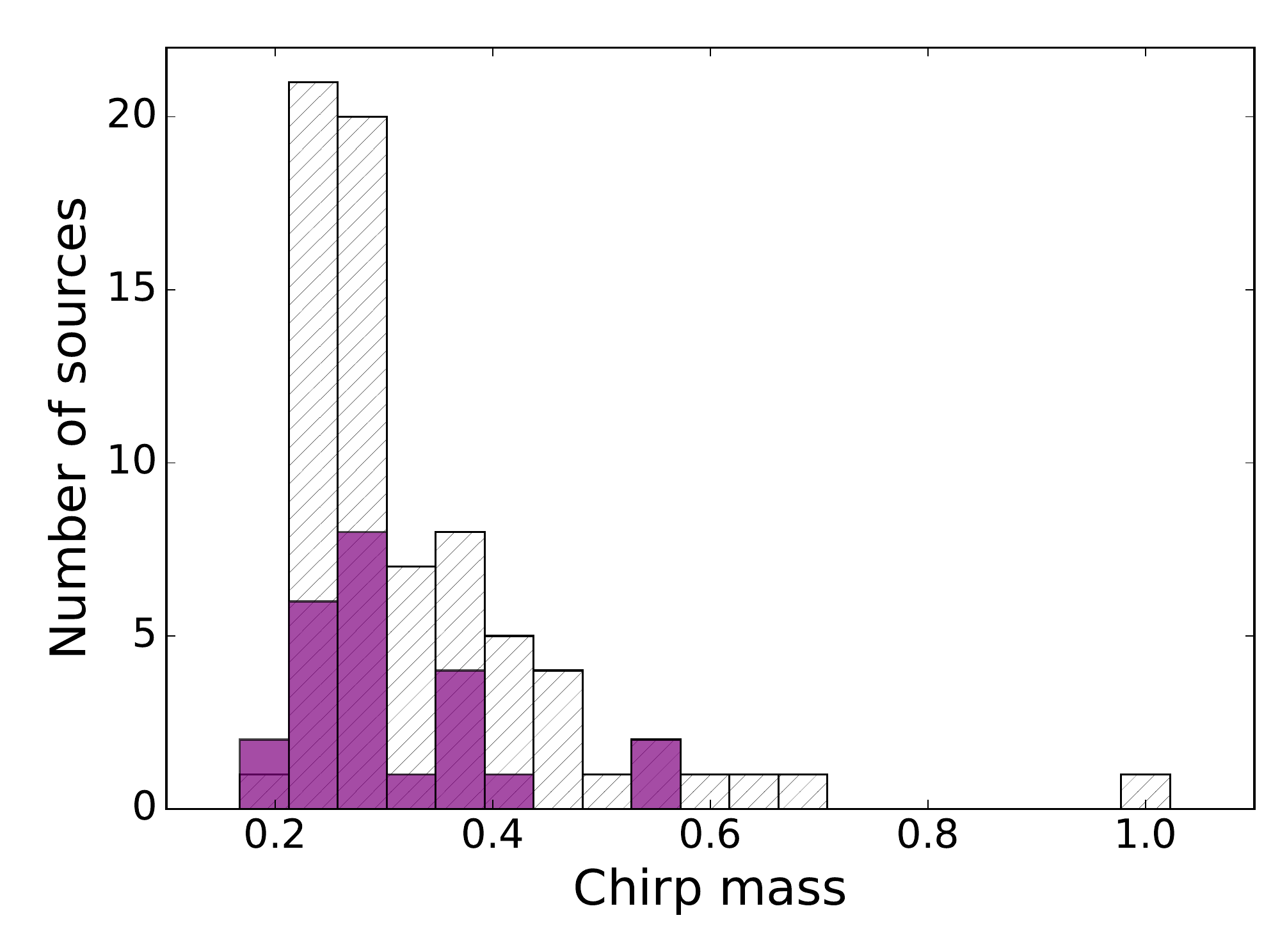}
	 \includegraphics[width=0.4\textwidth]{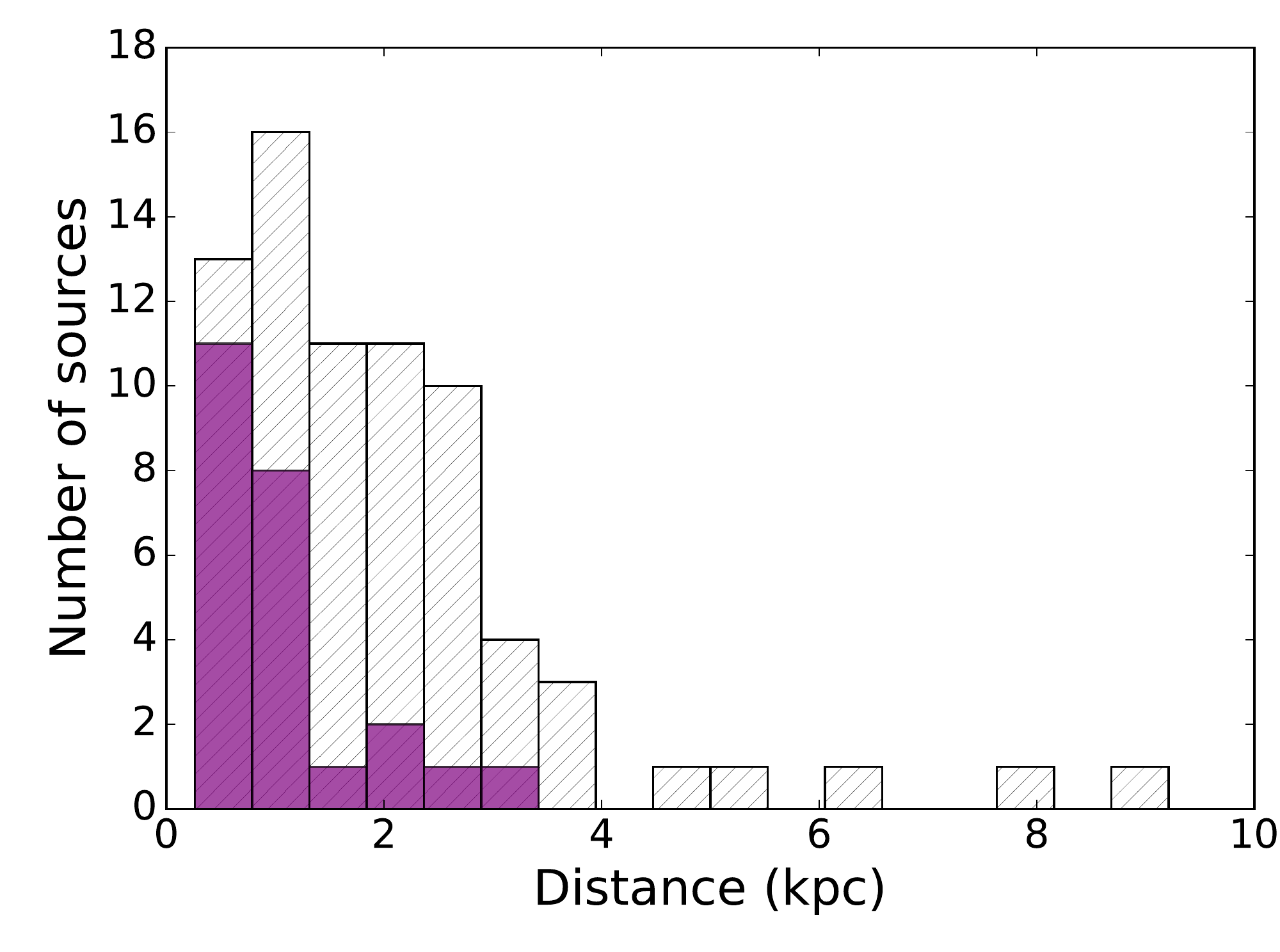}
         \caption{Number of combined EM and GW detections as a function of the chirp mass, binary orbital period and distance from the Sun. The red histogram represents Gaia - LISA and hatched histogram represents LSST - LISA combined detections.}
       \label{fig:10}
\end{figure}
%%%%%%%%%%%%%%%%%%%%%%%%%%%%%%

\section{Conclusions}

In this work we have computed the expected number of DWD detections by Gaia and the LSST as eclipsing sources, and by the future LISA mission as GW sources.
As in earlier studies we relied on population synthesis modeling because of the small number of the known systems (as in the case of detached DWDs considered here).
To simulate the Galactic population of DWD binaries we considered two different prescriptions for the CE phase ($\alpha \alpha$ and $\gamma \alpha$) in order to investigate whether Gaia, LSST and LISA will elucidate on the nature of the CE phase.
We find that Gaia can provide up to a few hundred of eclipsing detached DWDs, while LSST will extend this sample up to almost $2 \times 10^3$ sources.
We then investigated the number of individually resolvable GW sources in our model populations considering the latest mission concept of the LISA detector submitted as a response to the ESA call for L3 missions in 2017.
We find that the number of detectable detached DWDs is around $25 \times 10^3$ for the nominal 4 years of mission life time.
Finally, we used the obtained EM samples to estimate how many verification binaries Gaia and LSST will provide before the LISA launch.
We find several tens of combined EM and GW detection.
These detections will significantly increase the sample of know LISA verification binaries by at least factor of 2.

The subset of Gaia and LSST binaries analyzed in this work represent guaranteed detections for the LISA mission, and will provide a powerful tool for probing WD astrophysics and a unique opportunity of multi-messenger study for this class of objects.
No other GW sources are expected to provide so large number of combined GW and EM detections.
We defer to a future work the parameter estimation from EM and GW data for the sample of the combined EM and GW detections and the study of the applicability of these data to the study of the effects of tides in ultra-compact binaries and the kinematics of the Galaxy.

\section*{Acknowledgements}
We are grateful to Enrico Barausse, Antoine Petiteau, Alberto Sesana, Samaya Nissanke, Tom Marsh and Luke T. Maud for useful discussions.
This work was supported by the NWO WRAP Program.
\addcontentsline{toc}{section}{Acknowledgements}

\bibliographystyle{mnras}
%\bibliography{biblio} 

\appendix
\section{Appendix A - GW signal estimation}

To compute the LISA response for so a large amount of sources, as in our case, we follow a frequency based approach developed by \citet{CL2003}.
We report here the formulae from the original paper used in our simulation.

In the low frequency limit ($f \ll c/2\pi L$) the antenna pattern is well approximated by a quadrupole, so the GW signal as measured  by the detector can be expressed as
\begin{equation}
s(t) = F_+ h_+(t) +F_{\times} h_{\times}(t),
\end{equation}
where $F_+$ and $F_{\times}$ are the antenna pattern factors, that depend only on the angular coordinates of the source 
in the detector's reference frame $(\phi_{\rm d},\theta_{\rm d})$
and on the polarization angle $\psi_{\rm d}$\footnote{$\psi_{\rm d}$ is measured clockwise about ${\bf \hat{n}}$ from the orbit's line of nodes to the ${\bf \hat{x}}$-axis in the source reference frame.} of the wavefront.
For a two-arm Michelson-like interferometer $F_+$ and $F_{\times}$ can be written as
\begin{equation}
	\begin{split}
F_+ (\theta_{\rm d}, \phi_{\rm d}, \psi_{\rm d}) = & \frac{1}{2} (1 + \cos^2 \theta_{\rm d}) \cos 2\phi_{\rm d} \cos 2\psi_{\rm d} \\ 
& - \cos \theta_{\rm d} \sin 2\phi_{\rm d} \sin 2 \psi_{\rm d}) ,
	\end{split}
\end{equation}
\begin{equation}
	\begin{split}
F_{\times} (\theta_{\rm d}, \phi_{\rm d}, \psi_{\rm d}) = & \frac{1}{2} (1 + \cos^2 \theta_{\rm d}) \cos 2\phi_{\rm d} \sin 2\psi_{\rm d} \\
& + \cos \theta_{\rm d} \sin 2\phi_{\rm d} \cos 2 \psi_{\rm d}).
	\end{split}
\end{equation}

For a space-based mission, the orbital motion of the detector modulates the signal with time in amplitude, frequency and phase \citep[e.g.][]{Cutler1998,CL2003}.
The amplitude modulation is due to the fact that the sensitivity of the instrument is non uniform with the source position on the sky with respect to the detector.
The frequency modulation originates from the Doppler shift due to the relative motion of the detector with respect to the source.
Lastly, the phase modulation is due to the different response of the instrument to the two wave polarizations.
Thus, to take into account these effects for signals coming from all possible directions in the sky, it is convenient to refer all quantities to the heliocentric ecliptic frame, so that  the angular coordinate of the source are fixed in time, and the time modulation of the signal is given by the change of orientation of the detector with respect to the Sun \citep{Cutler1998,CR2003, CL2003}.
To keep the notation uniform with previous works in the following we indicate ecliptic coordinates $(\lambda, \beta)$ as $(\phi, \theta)$, where $\phi = \lambda$ and $\theta = \pi/2 - \beta$, so that $\phi \in [0,2 \pi]$ and $\theta \in [0, \pi]$. 

The signal of the source as seen by the detector can be expressed in amplitude-and-phase form as
\begin{equation}
s (t) = {\cal A}(t) \cos \left[ 2 \pi f t + \phi_{\rm P}(t) + \phi_{\rm D}(t) + \phi_0 \right],
\end{equation}
where
\begin{equation}
 {\cal A}(t) = {\left[ h_+^2 F_+^2 (t) + h_{\times}^2 F_{\times}^2 (t)\right]}^{1/2}
\end{equation}
\begin{equation}
\phi_{\rm P} (t) = - \tan^{-1} \left( \frac{h_{\times} F_{\times}(t)}{ h_+ F_+(t)} \right)
\end{equation}
\begin{equation}
\phi_{\rm D} = 2 \pi f c^{-1} R \sin \theta \cos ( 2 \pi f_{\rm m} t - \phi) 
\end{equation}
where ${\cal A}(t)$ is the wavefront amplitude, $\phi_{\rm P}$ is the polarization phase induced by the change of the LISA orientation, $\phi_{\rm D}$ is the difference between the phase of the wavefront at the detector and the phase of the wavefront at the Sun and $R=1$AU, $f_{\rm m} =1/$yr and $\phi_0$ is the initial GW phase.
Note, that now the time dependence of ${\cal A}(t)$ and $\phi_{\rm P} (t)$ is determined by $F_{\times}$ and $F_+$. 
Equations (A2)-(A3) can be written for the LISA (three-arm configuration) as
\begin{equation}
	F^+(t) = \frac{1}{2} \left[ \cos 2 \psi D^+(t) - \sin 2\psi D^{\times}(t) \right],
\end{equation}
\begin{equation}
	F^{ \times} (t) = \frac{1}{2} \left[ \sin 2 \psi D^+(t) + \cos 2\psi D^{\times}(t) \right],
\end{equation}
with $D^+$ and $D^{\times}$ given in terms of ecliptic coordinates $(\theta, \phi)$ defined above:
\begin{equation}
	\begin{split}
		D^+ (t) = \ & \frac{ \sqrt{3} }{64} \{  -36 \sin^2 \theta \sin (2 \alpha(t) -2 \lambda) + (3 + \cos 2\theta) \\ 
		& \times \{\cos 2\phi [9 \sin 2\lambda -\sin (4 \alpha(t) - 2 \lambda)] \\
		& + \sin 2\phi [  \cos(4 \alpha -2\lambda) -9 \cos2\lambda ] \}  \\
		& - 4 \sqrt{3} \sin 2\theta[ \sin(3\alpha(t) - 2\lambda -\phi ) \\
		& -3 \sin(\alpha(t) - 2\lambda + \phi)  ] \},
		\end{split}
\end{equation}
\begin{equation}
	\begin{split}
		D^{\times} (t) = \ & \frac{1}{16} \{ \sqrt{3} \cos \theta [ 9\cos(2 \lambda -2\phi) - \cos(4\alpha(t) - \\
		&2\lambda -2\phi)]  -6 \sin \theta [ \cos(3\alpha(t) -2 \lambda - \phi) \\
		&+ 3 \cos(\alpha(t) -2\lambda + \phi) ] \}. 
	\end{split}
\end{equation}
where $\alpha (t) = 2 \pi f_{\rm m} t + k$ is the orbital phase of the detector's barycenter,  $k$ and $\lambda$ specify the initial position and orientation of the detector, that as in \cite{CL2003} we set respectively to $0$ and $3 \pi / 4$.  

The orbit average amplitude of the signal produced at the detector can be found from eq. (A5) as
\begin{equation}
 \langle {\cal A} \rangle ^2 = \frac{1}{2} A^2 \left[ (1 + \cos^2 i)^2  \langle F_+^2 \rangle + 4 \cos^2 i \langle F_{\times}^2\rangle \right],
\end{equation}
where ${A = 2(G{\cal M})^{5/3}(\pi f)^{2/3}/(c^4d)}$ is the intrinsic amplitude of the waveform (see eqns.(7)-(8)), and the orbit average $F_+$ and $F_{\times}$ are given by
\begin{equation}
\begin{aligned}
 \langle F_+^2 \rangle &= \frac{1}{4} \left( \cos^2 2\psi  \langle D_+^2 \rangle - \sin 4 \psi  \langle D_+ D_{\times} \rangle
+ \sin^2 2\psi  \langle D_+^2 \rangle\right), \\
 \langle F_{\times}^2 \rangle &= \frac{1}{4} \left( \cos^2 2\psi  \langle D_{\times}^2 \rangle + \sin 4 \psi  \langle D_+ D_{\times} \rangle
+ \sin^2 2\psi  \langle D_+^2 \rangle\right),
\end{aligned}
\end{equation}
and 
\begin{equation}
\begin{aligned}
 \langle D_+ D_{\times} \rangle = &\frac{243}{512} \cos \theta \sin 2 \phi(2 \cos^2 \phi -1)(1+\cos^2 \theta), \\
 \langle D_{\times}^2 \rangle = &\frac{3}{512} (120 \sin^2 \theta + \cos^2 \theta + 162 \sin^2 2\phi \cos^2 \theta), \\
 \langle D_+^2 \rangle = &\frac{3}{2048} [ 487 + 158 \cos^2 + 7 \cos^4 \theta  \\
&- 162 \sin^2 2\phi (1+\cos^2 \theta)^2 ].
\end{aligned}
\end{equation}
We used the orbit averaged amplitude to compute the SNR (eq.(10)).
Note, that one should substitute in eq.(10) the non-sky-averaged noise spectral $S_{\rm n}^{\rm NSA}$, that can be obtained as $S_{\rm n}^{\rm NSA} (f) = 3/20 S_{\rm n}^{\rm SA} (f)$ from the sky-averaged (but not inclination averaged) spectral density \citep[][Sect.II.C]{Berti2005} represented in Fig. 10.

\section{Appendix B - List of the LISA verification binaries}

Table B.1 shows the sample of currently known DWD and AM CVn systems with expected SNR in GWs > 0.01, evaluated by using eq.(10) and the LISA ESACall v1.1 configuration sensitivity for 1 yr observation time.

%%%%%%%%%%%%%%%%%%%%%%%%%%%%%%
\begin{table*} 
\centering
\caption{A sample of known interacting (AM CVn stars) and non interacting (detached DWDs) LISA verification binaries. Amplitudes  are given in units of $10^{-23}$.  To compute the SNR for each binary we set the initial orbital phase and polarization angle to 0$^{\circ}$, and the inclination to 60$^{\circ}$ for cases where it is unknown.}
\label{tab:A1}
\begin{threeparttable}
\begin{tabular}{l |c|c|c|c|c|c|c|c|c|c}
\hline
\hline
Name         & $l_{\rm gal}$(deg) &$ b_{\rm gal}$(deg) & $P$(s)      & $m_1$(M$\odot$)  & $m_2$(M$\odot$)  & d(pc)  & $i$(deg) & $f_{GW}$(mHz) & ${\cal A} $  & SNR    \\ \hline 
RX J0806\tnote{a}   & 206.93      & 23.4        & 321.52911 & 0.55  & 0.27  & 5000.0 & 37.0   & 6.22       & 6.43  & 108.82 \\
V407 Vul \tnote{a}   & 57.73       & 6.44        & 569.395   & 0.6   & 0.07  & 2000.0 & 60.0   & 3.51       & 3.32   & 20.98  \\
ES Cet\tnote{a}   & 168.97      & -65.86      & 621.0     & 0.6   & 0.06  & 1000.0 & 60.0   & 3.22       & 5.4    & 23.72  \\
AM CVn\tnote{a}    & 140.23      & 78.94       & 1028.73   & 0.71  & 0.13  & 600.0  & 43.0   & 1.94       & 15.22  & 17.03  \\
SDSS J1908+3940\tnote{a}  & 70.66       & 13.93       & 1092.0    & 0.6   & 0.05  & 1000.0 & 60.0   & 1.83       & 3.11   & 2.34   \\
HP Lib\tnote{a}    & 352.06      & 32.55       & 1103.0    & 0.57  & 0.06  & 200.0  & 30.0   & 1.81       & 17.77  & 19.7   \\
PTF1J1919+4815\tnote{a}     & 79.59       & 15.59       & 1350.0    & 0.6   & 0.04  & 2000.0 & 60.0   & 1.48       & 1.08   & 0.56   \\
CR Boo\tnote{a}       & 340.96      & 66.49       & 1471.0    & 0.79  & 0.06  & 340.0  & 30.0   & 1.36       & 10.82  & 7.47   \\
KL Dra\tnote{a}       & 91.01       & 19.2        & 1500.0    & 0.6   & 0.02  & 1000.0 & 60.0   & 1.33       & 1.02   & 0.44   \\
V803 Cen\tnote{a}     & 309.37      & 20.73       & 1596.0    & 0.84  & 0.08  & 350.0  & 14.0   & 1.25       & 13.75  & 9.08   \\
SDSS J0926\tnote{a}   & 187.51      & 46.01       & 1699.0    & 0.85  & 0.04  & 460.0  & 83.0   & 1.18       & 5.13   & 1.03   \\
CP Eri\tnote{a}       & 191.7       & -52.91      & 1701.0    & 0.6   & 0.02  & 700.0  & 60.0   & 1.18       & 1.34   & 0.45   \\
2003aw\tnote{a}        & 235.13      & 26.48       & 2028.0    & 0.6   & 0.02  & 700.0  & 60.0   & 0.99       & 1.19   & 0.3    \\
2QZ 1427 -01\tnote{a}   & 345.67      & 37.17       & 2194.0    & 0.6   & 0.015 & 700.0  & 60.0   & 0.91       & 0.85   & 0.19   \\
SDSS J1240\tnote{a}   & 297.57      & 60.77       & 2242.0    & 0.6   & 0.01  & 400.0  & 60.0   & 0.89       & 0.98   & 0.23   \\
SDSS J0804 \tnote{a}  & 205.94      & 23.37       & 2670.0    & 0.6   & 0.01  & 400.0  & 60.0   & 0.75       & 0.87   & 0.15   \\
SDSS J1411\tnote{a}   & 91.89       & 63.82       & 2760.0    & 0.6   & 0.01  & 400.0  & 60.0   & 0.72       & 0.85   & 0.14   \\
GP Com\tnote{a}       & 323.55      & 80.3        & 2794.0    & 0.6   & 0.01  & 80.0   & 60.0   & 0.72       & 4.24   & 0.7    \\
SDSS J0902\tnote{a}   & 184.42      & 41.32       & 2899.0    & 0.6   & 0.01  & 500.0  & 60.0   & 0.69       & 0.66   & 0.1    \\
SDSS J1552\tnote{a}   & 51.31       & 50.53       & 3376.3    & 0.6   & 0.01  & 500.0  & 60.0   & 0.59       & 0.6    & 0.07   \\
CE 315\tnote{a}      & 309.26      & 39.25       & 3906.0    & 0.6   & 0.006 & 77.0   & 60.0   & 0.51       & 2.12   & 0.19   \\
J0651+2844\tnote{a}   & 186.93      & 12.69       & 765.4     & 0.55  & 0.25  & 1000.0 & 86.9   & 2.61       & 16.84  & 19.67  \\
J0935+4411\tnote{a}   & 176.0796    & 47.3776     & 1188.0    & 0.32  & 0.14  & 660.0  & 60.0   & 1.68       & 7.46   & 4.63   \\
J0106-1000\tnote{a}   & 135.72      & -72.47      & 2346.0    & 0.43  & 0.17  & 2400.0 & 67.0   & 0.85       & 1.95   & 0.37   \\
J1630+4233\tnote{a}   & 67.076      & 43.3603     & 2390.0    & 0.31  & 0.52  & 830.0  & 60.0   & 0.84       & 11.0   & 2.26   \\
J1053+5200\tnote{a}   & 156.4       & 56.79       & 3680.0    & 0.2   & 0.26  & 1100.0 & 60.0   & 0.54       & 2.44   & 0.24   \\
J0923+3028\tnote{a}   & 195.82      & 44.78       & 3884.0    & 0.279 & 0.37  & 228.0  & 60.0   & 0.51       & 20.13  & 1.73   \\
J1436+5010\tnote{a}   & 89.01       & 59.46       & 3957.0    & 0.24  & 0.46  & 800.0  & 60.0   & 0.51       & 5.91   & 0.51   \\
WD 0957-666\tnote{a} & 287.14      & -9.46       & 5296.81   & 0.32  & 0.37  & 135.0  & 68.0   & 0.38       & 31.07  & 1.27   \\
J0755+4906\tnote{a}   & 169.76      & 30.42       & 5445.0    & 0.176 & 0.81  & 2620.0 & 60.0   & 0.37       & 1.68   & 0.08   \\
J0849+0445\tnote{a}   & 222.7       & 28.27       & 6800.0    & 0.176 & 0.65  & 1004.0 & 60.0   & 0.29       & 3.22   & 0.09   \\
J0022-1014\tnote{a}   & 99.2997     & -71.7538    & 6902.496  & 0.21  & 0.375 & 1151.0 & 60.0   & 0.29       & 2.15   & 0.06   \\
J2119-0018\tnote{a}   & 51.58       & -32.54      & 7497.0    & 0.74  & 0.158 & 2610.0 & 60.0   & 0.27       & 1.15   & 0.02   \\
J1234-0228\tnote{a}   & 294.25      & 60.11       & 7900.0    & 0.09  & 0.23  & 716.0  & 60.0   & 0.25       & 1.01   & 0.02   \\
WD 1101+364\tnote{a} & 184.48      & 65.62       & 12503.0   & 0.36  & 0.31  & 97.0   & 25.0   & 0.16       & 23.22  & 0.19   \\
WD 0931+444\tnote{b} & 176.08      & 47.38       & 1200.0    & 0.32  & 0.14  & 660.0  & 70.0   & 1.67       & 7.41   & 3.58   \\
WD 1242-105\tnote{c} & 300.31      & 51.98       & 10260.0   & 0.56  & 0.39  & 39.0   & 45.1   & 0.19       & 114.75 & 1.41   \\
J0056-0611\tnote{d}   & 126.6604    & -69.0278    & 3748.0    & 0.174 & 0.46  & 585.0  & 60.0   & 0.53       & 6.28   & 0.63   \\
J0106-1000\tnote{d}    & 135.7244    & -72.4861    & 2345.76   & 0.191 & 0.39  & 2691.0 & 60.0   & 0.85       & 1.79   & 0.39   \\
J0112+1835\tnote{d}    & 129.77      & -44.0119    & 12699.072 & 0.62  & 0.16  & 662.0  & 60.0   & 0.16       & 2.84   & 0.01   \\
J0345+1748\tnote{d}    & 171.051     & -28.4018    & 20306.592 & 0.76  & 0.181 & 166.0  & 60.0   & 0.1        & 10.81  & 0.01   \\
J0745+1949\tnote{d}    & 200.4746    & 20.4396     & 9711.36   & 0.1   & 0.156 & 270.0  & 60.0   & 0.21       & 1.9    & 0.02   \\
J0751-0141\tnote{d}    & 221.4565    & 12.5761     & 6912.864  & 0.97  & 0.194 & 1859.0 & 60.0   & 0.29       & 2.52   & 0.07   \\
J0825+1152\tnote{d}    & 212.5705    & 26.1227     & 5027.616  & 0.49  & 0.287 & 1769.0 & 60.0   & 0.4        & 2.8    & 0.14   \\
J1053+5200\tnote{d}    & 156.4021    & 56.794      & 3677.184  & 0.26  & 0.213 & 1204.0 & 60.0   & 0.54       & 2.36   & 0.23   \\
J1054-2121\tnote{d}    & 269.7458    & 33.8695     & 9019.296  & 0.39  & 0.168 & 751.0  & 60.0   & 0.22       & 2.33   & 0.03   \\
J1056+6536\tnote{d}    & 140.067     & 47.5033     & 3759.264  & 0.34  & 0.338 & 1421.0 & 60.0   & 0.53       & 3.62   & 0.35   \\
J1108+1512\tnote{d}    & 234.1026    & 63.2376     & 10635.84  & 0.42  & 0.167 & 698.0  & 60.0   & 0.19       & 2.36   & 0.02   \\
J1112+1117\tnote{d}    & 242.321     & 61.8382     & 14902.272 & 0.14  & 0.169 & 257.0  & 60.0   & 0.13       & 2.14   & 0.01   \\
J1130+3855\tnote{d}    & 172.9043    & 69.3762     & 13523.328 & 0.72  & 0.286 & 662.0  & 60.0   & 0.15       & 5.2    & 0.02   \\
J1436+5010\tnote{d}    & 89.0112     & 59.4607     & 3957.12   & 0.46  & 0.233 & 830.0  & 60.0   & 0.51       & 5.55   & 0.48   \\
J1443+1509\tnote{d}    & 14.0206     & 61.3102     & 16461.792 & 0.84  & 0.181 & 540.0  & 60.0   & 0.12       & 4.11   & 0.01   \\
J1630+4233\tnote{d}    & 67.076      & 43.3603     & 2389.824  & 0.3   & 0.307 & 820.0  & 60.0   & 0.84       & 7.06   & 1.45   \\
J1741+6526\tnote{d}    & 95.1544     & 31.7085     & 5279.904  & 1.11  & 0.17  & 936.0  & 60.0   & 0.38       & 5.82   & 0.28   \\
J1840+6423\tnote{d}    & 94.3694     & 25.424      & 16528.32  & 0.65  & 0.177 & 676.0  & 60.0   & 0.12       & 2.66   & 0.01   \\
J2338-2052\tnote{d}    & 49.5602     & -72.1995    & 6604.416  & 0.15  & 0.263 & 1295.0 & 60.0   & 0.3        & 1.11   & 0.03   \\
CSS 41177\tnote{e}   & 210.129     & 52.424      & 8208.0    & 0.36  & 0.31  & 473.0  & 88.9   & 0.24       & 6.3    & 0.06   \\
J1152+0248\tnote{f}   & 270.23      & 61.86       & 8602.0    & 0.47  & 0.41  & 464.0  & 89.2   & 0.23       & 9.82   & 0.1 \\  
\hline
\end{tabular}
\begin{tablenotes}[para] 
\item[a] http://www.astro.ru.nl/~nelemans/dokuwiki/doku.php?id=verification\_binaries:intro, 
\item[b] \citet{Kilic2014},
\item[c] \citet{Debes2015},
\item[d]\citet{Gianninas2015},
\item[e]\citet{Bours2014},
\item[f]\citet{Hallakoun2016}.
\end{tablenotes}
\end{threeparttable}
\end{table*}

\end{document}